\documentclass[journal=jacsat,manuscript=article]{achemso}
\usepackage{graphicx}
\usepackage[T1]{fontenc} 
\usepackage[dvipsnames]{xcolor}

\author{Mathieu Massicotte}
 \affiliation{Institut quantique, Département de physique, Université de Sherbrooke, Sherbrooke, QC, J1K 2R1, Canada}
 \affiliation{Institut Interdisciplinaire d’Innovation Technologique, Laboratoire Nanotechnologies Nanosystèmes – CNRS, Département de génie électrique et génie informatique, Université de Sherbrooke, Sherbrooke, J1K 2R1, Canada}
 \email{mathieu.massicotte@usherbrooke.ca}
 \author{Sam Dehlavi}
 \affiliation{Institut quantique, Département de physique, Université de Sherbrooke, Sherbrooke, QC, J1K 2R1, Canada}
 \author{Xiaoyu Liu}
 \affiliation{Department of Physics, University of Toronto, Toronto, Ontario, M5S 1A7, Canada}
 \author{James L. Hart}
 \affiliation{Department of Materials Science and Engineering, Cornell University, Ithaca, New York 14853, USA} 
 \author{Elio Garnaoui}
 \affiliation{Institut quantique, Département de physique, Université de Sherbrooke, Sherbrooke, QC, J1K 2R1, Canada}
\author{Paula Lampen-Kelley}
\author{Jiaqiang Yan}
\author{David Mandrus}
\affiliation{Materials Science and Technology Division, Oak Ridge National Laboratory, Oak Ridge, TN 37831, USA and Department of Materials Science and Engineering, University of Tennessee, Knoxville, TN 37996}
\author{Stephen E. Nagler}
\affiliation{Neutron Scattering Division,  Oak Ridge National Laboratory, Oak Ridge, Tennessee 37831, USA and
 Department of Physics and Astronomy, The University of Tennessee, Knoxville, Tennessee 37996, USA}
\author{Kenji Watanabe}
\author{Takashi Taniguchi}
\affiliation{National Institute for Materials Science, 1-1 Namiki, Tsukuba 305-0044, Japan}
\author{Bertrand Reulet}
 \affiliation{Institut quantique, Département de physique, Université de Sherbrooke, Sherbrooke, QC, J1K 2R1, Canada}
\author{Judy J. Cha}
\affiliation{Department of Materials Science and Engineering, Cornell University, Ithaca, New York 14853, USA} 
\author{Hae-Young Kee}
 \affiliation{Department of Physics, University of Toronto, Toronto, Ontario, M5S 1A7, Canada} 
\author{Jeffrey A. Quilliam}
 \affiliation{Institut quantique, Département de physique, Université de Sherbrooke, Sherbrooke, QC, J1K 2R1, Canada}

\title{Giant anisotropic magnetoresistance in few-layer $\alpha$-RuCl$_3$ tunnel junctions}

\usepackage{atbegshi}
\AtBeginDocument{\AtBeginShipoutNext{\AtBeginShipoutDiscard}}

\begin{document}
\pagenumbering{arabic}

\newpage

\begin{abstract}
The spin-orbit assisted Mott insulator $\alpha$-RuCl$_3$ is proximate to the coveted quantum spin liquid (QSL) predicted by the Kitaev model. In the search for the pure Kitaev QSL, reducing the dimensionality of this frustrated magnet by exfoliation has been proposed as a way to enhance magnetic fluctuations and Kitaev interactions. Here, we perform angle-dependent tunneling magnetoresistance (TMR) measurements on ultrathin $\alpha$-RuCl$_3$ crystals with various layer numbers to probe their magnetic, electronic and crystal structure. We observe a giant change in resistance --  as large as $\sim$2500$\%$  -- when the magnetic field rotates either within or out of the $\alpha$-RuCl$_3$ plane, a manifestation of the strongly anisotropic spin interactions in this material. In combination with scanning transmission electron microscopy, this tunneling anisotropic magnetoresistance (TAMR) reveals that few-layer $\alpha$-RuCl$_3$ crystals remain in the high-temperature monoclinic phase at low temperature. It also shows the presence of a zigzag antiferromagnetic order below the critical temperature $T_N \simeq 14$ K, which is twice the one typically observed in bulk samples with rhombohedral stacking. Our work offers valuable insights into the relation between the stacking order and magnetic properties of this material, which helps lay the groundwork for creating and electrically probing exotic magnetic phases like QSLs via van der Waals engineering.
\end{abstract}

\section{Introduction}

The quantum spin liquid (QSL) is an elusive state of matter in which quantum fluctuations and magnetic frustration generate long-range quantum entanglement and prevent magnetic ordering down to zero temperature~\cite{Savary2017,Anderson1973,Balents2010}. One prominent type of QSL, predicted by the exactly solvable Kitaev model~\cite{Kitaev2006}, has two varieties of fractional excitation, Majorana fermions and fluxes, as its elementary excitations. The novel quantum statistics of these excitations make it promising for topological quantum computation~\cite{Nayak2008}. It was later shown that this model could be materialized in spin-orbit assisted Mott insulators with bond-dependent Ising interactions~\cite{Jackeli2009,Takagi2019}. The ensuing search for candidate Kitaev materials led to the emergence of $\alpha$-RuCl$_3$, a van der Waals material, as one of the main prospects~\cite{Plumb2014,Johnson2015,Sears2015,Majumder2015}. Despite early studies reporting an unconventional continuum of magnetic excitations~\cite{Banerjee2017,Banerjee2016,Nasu2016}, $\alpha$-RuCl$_3$ presents a zigzag antiferromagnetic (AFM) ground state, which preempts the realization of a pure QSL ground state. Nevertheless, this magnetic order can be quenched by applying an in-plane magnetic field larger than a critical field $B_c$, typically on the order of 6 - 8 T~\cite{Zheng2017,Sears2017,Wolter2017,Wang2017,Baek2017}. The report of half-integer quantization of the thermal Hall conductance just above this critical field generated huge interest, as it provided strong evidence of a Kitaev QSL phase in this field regime~\cite{Kasahara2018}. Yet, because of reproducibility issues caused by sample variations~\cite{Kim2022, Yamashita2020, Lefrancois2022}, other approaches are currently being explored to suppress magnetic order and enhance Kitaev interaction in $\alpha$-RuCl$_3$.

A tantalizing route to realizing a true Kitaev QSL in $\alpha$-RuCl$_3$ is to reduce its dimensionality via mechanical exfoliation in order to enhance order parameter fluctuations. Raman spectroscopy studies reported robust magnetic fluctuations down to the monolayer as well as the presence of lattice distortions~\cite{Du2019,Lin2020,Zhou2019}, which were also observed by low energy electron diffraction measurements~\cite{Dai2020}. Such mechanical strain may alter the magnetic phase of the monolayer, as suggested by {\it ab initio} calculations~\cite{Vatansever2019}. Recent tunneling studies also reported these distortions~\cite{Zheng2023} and found that they can lead to a reversal of the magnetic anisotropy to easy-axis anisotropy for the monolayer~\cite{Yang2023a}. The van der Waals nature of $\alpha$-RuCl$_3$ also opens the door to manipulating and probing its magnetic state by coupling it to other two-dimensional (2D) materials. Electronic transport and optical studies on $\alpha$-RuCl$_3$/graphene heterostructures reported a large charge transfer between the two layers~\cite{Wang2020, Mashhadi2019, Rizzo2020, Zhou2019a}, in agreement with first-principles calculations~\cite{Souza2022,Biswas2019,Gerber2020}. Transport measurements on these heterostructures also revealed proximity effects at low temperature, hinting at the presence of an ordered magnetic ground state in exfoliated $\alpha$-RuCl$_3$ crystals~\cite{Mashhadi2019,Zhou2019}. However, the magnetic and crystal structure of these flakes remain mostly unknown. In particular, the relation between the stacking order and magnetic properties of this material remains poorly understood, even for bulk crystals~\cite{Kim2022}.

\begin{figure*}[hbtp]
	\includegraphics[width=\textwidth]{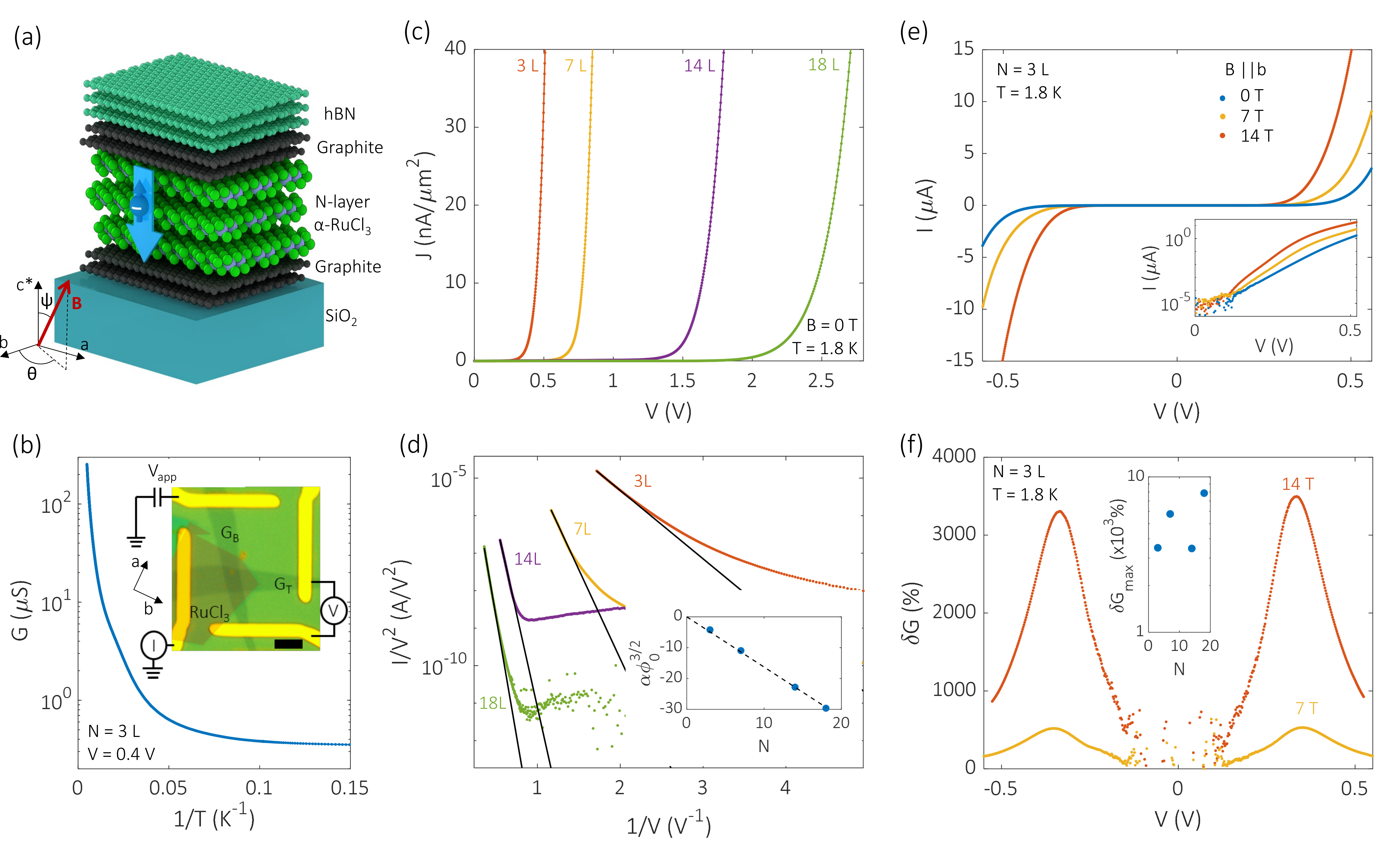}
	\caption{{\bf Graphite/$\alpha$-RuCl$_3$/graphite MTJs.} {\bf a} Schematic representation of electron tunneling in a van der Waals heterostructure under an external magnetic field $B$ (red arrow). The heterostructure comprises two graphite sheets separated by $\alpha$-RuCl$_3$ barrier with $N = 3$ layers. It is covered by a flake of hBN and deposited on SiO$_2$/Si substrate. The direction of the $B$ field is defined by the azimuthal angle $\psi$ and polar angle $\theta$. {\bf b} Temperature dependence of conductance $G$ of the trilayer device ($N = 3$) at zero field and $V = 0.4$ V. Inset: Optical image of a trilayer MTJ. The $\alpha$-RuCl$_3$ is colored in orange for clarity. The top (G$_T$) and bottom (G$_B$) graphite are deposited on top of Au/Ti contacts. $V_{app}$ represents the voltage applied to the lead while $V$ is the voltage measured across the junction. The scale bar is 10 $\mu$m. {\bf c} Current density $J = I/A$ ($A$ is the area of the junction) as a function of $V$ for MTJs with various numbers of $\alpha$-RuCl$_3$ layer ($N$) at zero field and 1.8 K. {\bf d} Plot of $\ln(I/V^2)$ as a function of $1/V$ for the same measurement as in {\bf c}. The black lines are linear fits corresponding to the FN tunneling model. Inset:  $\alpha \phi_0^{3/2}$ vs $N$. The blue points are obtained from the fits shown in the main panel while the dashed line is a linear fit to the data. {\bf e} Current $I$ as a function of $V$ measured at 1.8 K in a trilayer device under various values of $B$ field pointing along the $b$ axis (see inset of {\bf b}). Inset: Same data in a semi-log scale. {\bf f} Magnetoconductance $\delta G$ of the trilayer device as a function of $V$ for $B = 7$ and 14 T. The maximum indicates the onset of the FN tunneling regime.  Inset: The maximum magnetoconductance measured at $B = 14$ T in devices with different layer numbers $N$. }
	\label{Figure1}
\end{figure*}

Here, we investigate the magnetic and crystal structure of few-layer $\alpha$-RuCl$_3$ flakes by measuring their angle-dependent tunneling magnetoresistance (TMR) (Fig.~\ref{Figure1}a). This device-oriented technique, which has proven successful in studying other 2D magnets~\cite{Yang2023a,Wang2019,Klein2018,Wang2021,Song2018,Wang2018,Cai2019,Long2020,Kim2019a,Kim2018,Kim2019,Klein2019}, provides a sensitive and versatile tool to probe magnetism in nanoscale materials. Unlike other 2D materials, we observe a giant anisotropy of the TMR in $\alpha$-RuCl$_3$. This effect enables us to track the magnetic phase diagram and magnetocrystalline anisotropy of $\alpha$-RuCl$_3$ flakes with thickness ranging from 3 to 18 layers. Our results indicate that they host an AFM ground state with enhanced critical field and temperature compared to most bulk samples, and have a monoclinic crystal structure, which is supported by a high-resolution scanning transmission microscopy (STEM) study of isolated flakes. We use this knowledge to calculate the electronic structure of few-layer $\alpha$-RuCl$_3$ and quantitively explain our TMR measurements.  

\section{Results}

\subsection{Vertical transport in graphite/$\alpha$-RuCl$_3$/graphite heterostructures} 

Our magnetic tunnel junctions (MTJ) are made by placing an exfoliated $\alpha$-RuCl$_3$ flake between two graphite flakes, capped by a crystal of hBN (see Methods and Supporting Information, Figure S1, for details about device fabrication). The graphite contacts are arranged in a cross geometry, allowing for precise four-probe measurements of the tunneling conductance $G = I/V$ that avoids unwanted contributions from the graphite magnetoresistance. The inset of Fig.~\ref{Figure1}b presents an optical image of a typical device comprising a trilayer $\alpha$-RuCl$_3$ flake. The conductance of every junction decreases in a thermally activated way as the junction is cooled down to a temperature $T$ of about 30 K (Fig.~\ref{Figure1}b). The value of this activation energy is almost identical for all junctions, regardless of the $\alpha$-RuCl$_3$ thickness, with an average of $\phi_0 = 0.26 \pm 0.03$ eV (see Supporting Information, Figure S2). We attribute this behavior to thermionic transport over the potential barrier of height $\phi_0$ formed at the graphite/$\alpha$-RuCl$_3$ interface. This value of $\phi_0$ is consistent with a recent scanning tunneling spectroscopy study on few-layer $\alpha$-RuCl$_3$/graphite which indicates that upper bands are $\sim 0.3$ eV above the Fermi level~\cite{Zheng2023}.

At low temperature, the conductivity shows a weak dependence on temperature, indicating that transport across the junction is dominated by electron tunneling. The strongly nonlinear $I$-$V$ curves measured at $T = 1.8$~K (Fig.~\ref{Figure1}c) imply that $\alpha$-RuCl$_3$ flakes behave as insulating barriers. At high bias, the transport can be described by the Fowler-Nordheim (FN) tunneling~\cite{Simmons1963} relation  
$ I \propto \frac{V^2}{\phi_0} e^{\alpha \phi_0^{3/2}/V} $, where $\alpha = \frac{4tN\sqrt{2m^\ast}}{3\hbar q} $
(see Fig.~\ref{Figure1}d). Here, $t = 0.6$ nm is the layer thickness of $\alpha$-RuCl$_3$, $N$ is the number of layers, $m^\ast$ is the effective mass of carriers along the tunneling direction, $\hbar$ is the reduced Planck constant and $q$ is the elementary charge. The inset of Fig.~\ref{Figure1}d shows a linear relation between $\alpha \phi_0^{3/2}$ and $N$ obtained from the FN fits. Using the value of $\phi_0$ deduced from the temperature dependence of $G$, we estimate the effective mass inside the $\alpha$-RuCl$_3$ barrier to be $m^\ast \simeq 9 m_0$, where $m_0$ is the free electron mass. This large $m^\ast$ value reflects the highly correlated nature of electrons and the resulting non-dispersive bands in $\alpha$-RuCl$_3$, as confirmed below by {\it ab initio} calculations.

Having established that electron tunneling through the Mott-insulating $\alpha$-RuCl$_3$ is the dominant transport mechanism at low temperature, we investigate the effect of applying an in-plane magnetic field $B$ on the transport properties. Figure 1e shows a large increase of the tunneling current in the trilayer junction, which implies that the tunneling probability is linked to the field-dependent magnetic structure of $\alpha$-RuCl$_3$. The junction magnetoconductance $\delta G = [G(B,V)-G(0,V)]/G(0,V)$, plotted in Figure~\ref{Figure1}f, reaches a peak at finite bias. As observed in other MTJs~\cite{Song2018,Klein2019}, this peak corresponds to the onset of the FN tunneling regime. As we demonstrate below, the magnetoconductance in this regime can be well described by a spin-dependent tunneling model~\cite{Wang2021,Kim2018,Esaki1967} where the majority-spin electrons experience a lower energy barrier than that of minority-spin electrons. As a result, our junctions display large ($> 3000 \%$) magnetoconductance under an in-plane magnetic field of 14~T. This large magnetoconductance provides a sensitive probe to study the magnetic structure of exfoliated $\alpha$-RuCl$_3$ flakes. Unless otherwise specified, all measurements presented below are obtained from the trilayer device. Other devices exhibit qualitatively similar behavior and are presented in the Supporting Information.

\begin{figure*}[hbtp]
	\includegraphics[width=\textwidth]{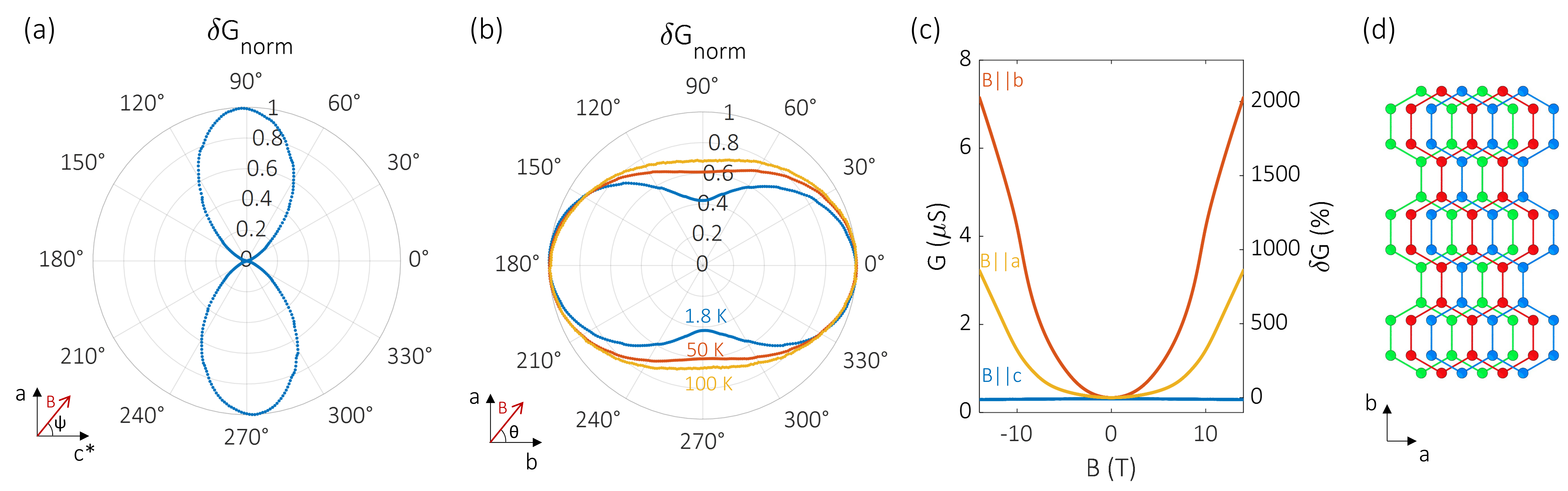}
	\caption{{\bf Angle-dependent magnetoconductance of a trilayer $\alpha$-RuCl$_3$ MTJ.} {\bf a} Polar plot of the normalized magnetoconductance $\delta G_\mathrm{norm}$ at  $B = 14$ T and $T = 1.8$ K as a function of $\psi$ in the $ac^\ast$ plane. {\bf b} Polar plot of the normalized magnetoconductance $\delta G_\mathrm{norm}$ at $B = 14$ T as a function of $\theta$ in the $ab$ plane at $T = $1.8, 50 and 100 K. {\bf c} Conductance $G$ (left $y$ axis) and magnetoconductance $\delta G$ (right $y$ axis) as a function of $B$ pointing along the three symmetry axes. These measurements, which were performed by sweeping $B$ in both directions, do not show any sign of hysteretic behavior.  All measurements in this figure are performed with $V_\mathrm{app} = 0.4$ V. {\bf d} Top view of the stacking order in the monoclinic C2/m phase for three layers. For simplicity, only the Ru atoms are represented as blue, red and green spheres, from the top to the bottom layer.}
	\label{Figure2}
\end{figure*}

\subsection{Magnetic anisotropy} 

First, we examine the magnetic anisotropy of $\alpha$-RuCl$_3$ crystals by measuring $G$ as a function of the orientation of $B$ with respect to the $c^\ast$ axis (angle $\psi$, Fig.~\ref{Figure2}a) and within the $ab$ plane (angle $\theta$, Fig 2b). While a large increase of the conductance $G_{ab}$ is observed when $B$ lies in the $ab$ plane, only a small decrease of $G_{c^\ast}$ ($\sim -1\%$) is detected when $B$ is perpendicular to the $\alpha$-RuCl$_3$ plane (Fig.~\ref{Figure2}c).  We attribute this negative $\delta G$ to the the positive magnetoresistance of the graphite contacts~\cite{Gopinadhan2015}. This leads to a giant tunneling anisotropic magnetoresistance effect, TAMR $= (G_{ab}  - G_{c^\ast} )/G_{c^\ast}$, as high as  $\sim2500\%$. This effect typically arises when electrons are tunneling into a material with large spin-orbit coupling and magnetic anisotropy \cite{Matos-Abiague2009,Tang2022}, as it is the case here. In $\alpha$-RuCl$_3$, the spin-orbit coupling leads to an off-diagonal AFM exchange interaction $\Gamma$ that is comparable in size to the Kitaev interaction~\cite{Sears2020,Rau2014,Chaloupka2016}. This $\Gamma$ interaction forces the moments to lie in the $ac$ plane with a finite $c$-axis component, which gives rise to the strong easy-plane magnetic anisotropy observed in bulk crystals~\cite{Johnson2015,Sears2015,Majumder2015,Sears2020,Kubota2015} and the large out-of-plane TAMR we measure. 

The in-plane magnetoconductance also displays significant anisotropy (Fig.~\ref{Figure2}b), resulting in an in-plane TAMR ratio, $(G_b-G_a)/G_a $, of up to $\sim 120\%$ at low temperature. The two-fold symmetry of the TAMR is observed over the entire range of magnetic field (see Fig.~\ref{Figure2}c and Supporting Information, Figure S3). It also survives at high temperature ($T > 100$ K), far above the magnetic transition temperature typically observed in $\alpha$-RuCl$_3$ ($T_N \simeq 7$ -14~K). This suggests that this in-plane TAMR does not stem from a long-range magnetic order but rather from the anisotropy of the spin Hamiltonian~\cite{Cen2022}. These observations contrast with the six-fold periodicity recently reported for bulk $\alpha$-RuCl$_3$~\cite{Tanaka2022,Balz2021}, but they match the two-fold symmetry of the in-plane susceptibility observed by Lampen-Kelley \emph{et al.}~\cite{Lampen-Kelley2018} in other bulk samples. The latter demonstrated that this effect is consistent with a monoclinic C2/m crystal structure (Fig.~\ref{Figure2}d), where the $a$ ($b$) axis corresponds to the direction of minimum (maximum) susceptibility. Accordingly, in our samples, we ascribe the direction of lower (higher) magnetoconductance to the $a$ ($b$) axis of the $\alpha$-RuCl$_3$ flake. We note that the extrema of $\delta G(\theta)$ in the $ab$ plane often coincide with the orientation of one of the crystal edges~\cite{Guo2016} of the $\alpha$-RuCl$_3$ flakes (e.g., see the inset of Fig.~\ref{Figure1}b), suggesting that the TAMR is indeed linked to the symmetry of the magnetocrystalline anisotropy. Finally, we  point out that similar out-of-plane and in-plane TAMR are observed in thicker flakes (see Supporting Information, Figure S4).

\subsection{Magnetic phase diagram}

\begin{figure*}[hbtp]
	\includegraphics[width=\textwidth]{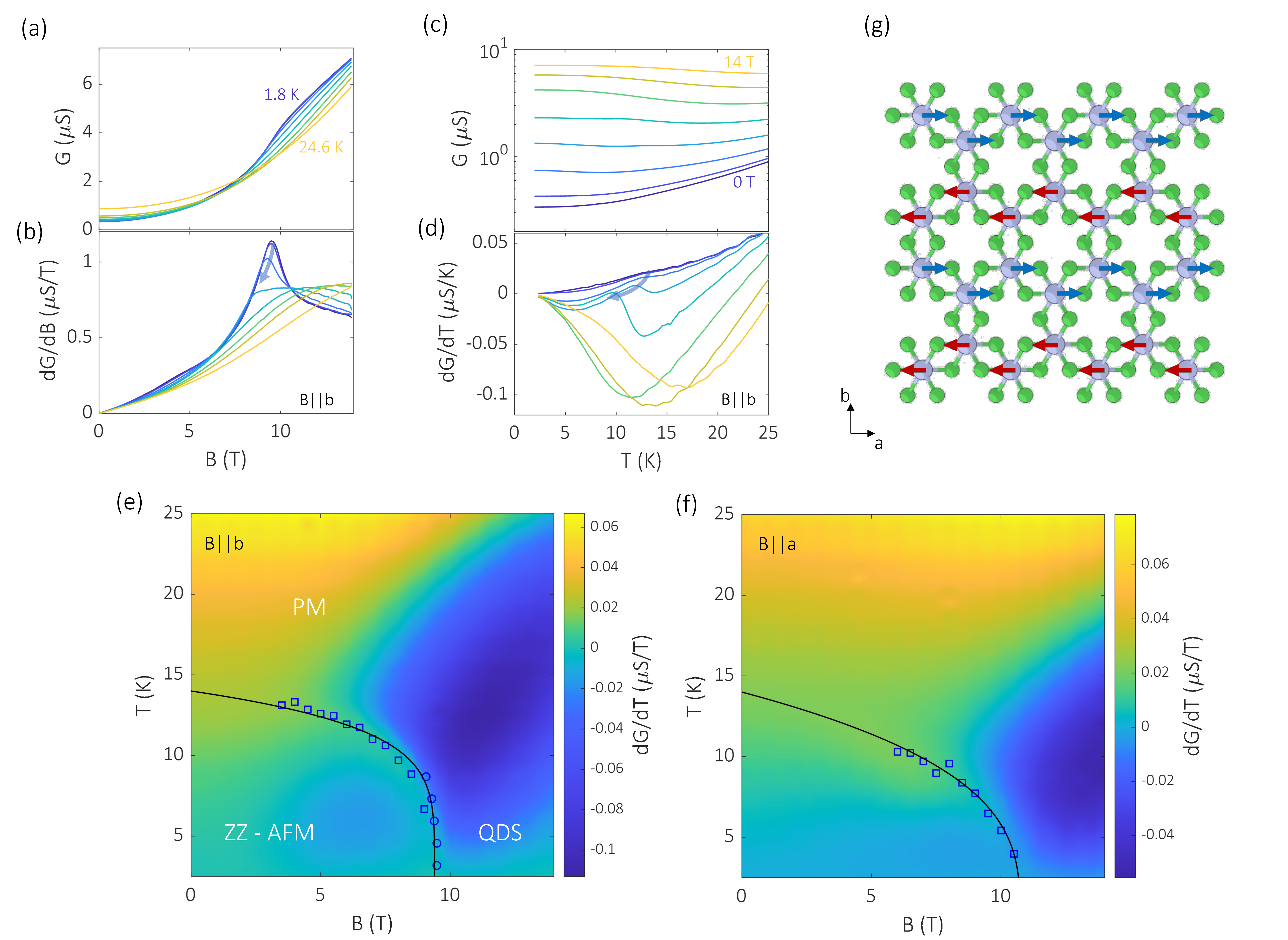}
	\caption{{\bf Magnetic phase diagram of trilayer $\alpha$-RuCl$_3$.}  {\bf a} Conductance $G$ and its derivative $dG/dB$ ({\bf b}) as a function of magnetic field $B$ at selected values of temperature $T$ between 1.8 and 24.6 K. {\bf c} Conductance $G$ and its derivative $dG/dT$ ({\bf d}) as a function of $T$ at selected values of $B$ between 0 and 14 T . The arrows in {\bf b} and {\bf d} indicate the position and evolution of peaks in the derivative. {\bf e,f} Color plot of $dG/dT$ as a function of $T$ and $B$, with $B$ pointing along the $b$ and $a$ axes of the crystal, respectively. Three phases are identified: zigzag antiferromagnetic (ZZ-AFM), paramagnetic (PM) and partially polarized quantum disordered state (QDS). The blue circles and squares correspond to the position of peaks in $dG/dB$ and $dG/dT$, respectively. The black lines are fits to those data points, as described in the main text. All measurements in this figure are performed with $V_\mathrm{app} = 0.4$~V. {\bf g} Top view of the crystal and ZZ-AFM spin structure of a $\alpha$-RuCl$_3$ monolayer. According to Ref.~\cite{Cao2016}, spins (red and blue arrows) lie in the $ac$ plan. Ru and Cl atoms are represented by gray and green spheres, respectively. }
	\label{Figure3}
\end{figure*}

In addition to identifying the magnetocrystalline axes of our $\alpha$-RuCl$_3$ flakes, magnetoconductance measurements allow us to probe their magnetic phase boundaries. We do so by measuring the conductance $G$ as a function of temperature $T$ and magnetic field $B$ applied along their $a$ and $b$ axes. Figure~\ref{Figure3}a shows $G$ as a function of $B$ at selected values of $T$ for $B||b$. At low temperature, $G$ exhibits a nearly quadratic field dependence which becomes almost linear when $B > B_C \simeq 9$ - 10 T. Beyond this critical field, $G$ does not appear to entirely saturate, even up to 14 T.  This transition is made more visible by taking the derivative of $G$ with respect to $B$ as shown in Figure~\ref{Figure3}b. We see that the peak corresponding to the transition shifts towards lower field and decreases as temperature increases. To further investigate this transition, we consider the temperature dependence of $G$ at selected values of $B$ (Fig.~\ref{Figure2}c) and its derivative $dG/dT$ (Fig.~\ref{Figure3}d).  At low field ($B < B_C$), a small peak is observed in $dG/dT$ at a Néel temperature $T_N \sim 14$ K, which shifts to lower temperatures as $B$ increases. In contrast, at high field ($B > B_C$), a dip appears in $dG/dT$ that moves to higher temperatures with increasing $B$.  

These results can be represented more effectively by plotting the colour map of $dG/dT$ as a function of $T$ and $B$, as illustrated in Figure~\ref{Figure3}e. This map is strongly reminiscent of a typical magnetic field-temperature phase diagram. We see that the phase boundary coincides well with the position of the peak extracted from the $dG/dB$ and $dG/dT$ curves (open blue circles and squares, respectively). 
This phase boundary can be fitted using the power law $T_N(B) = T_N(0)(1 - B/B_c)^{\nu z}$, where the exponent $\nu z \simeq 0.16$,  $B_c \simeq 9.4$ T and the zero-field Néel temperature $T_N(0) \simeq 14$ K.  We also performed similar measurements and analysis with $B||a$, as shown in Figure~\ref{Figure3}f.  In this case, the phase boundary is best described by the power law parameters $\nu z \simeq 0.33$ and $B_c \simeq 10.7$ T. We note that a similar phase boundary is observed in thicker $\alpha$-RuCl$_3$ flakes (see Supporting Information, Figure S5).

\section{Discussion}

The origin of this phase boundary can be readily interpreted by comparing it to the one observed in bulk $\alpha$-RuCl$_3$~\cite{Johnson2015,Zheng2017,Sears2017,Tanaka2022,Balz2021}. Like in the bulk case, we ascribe the magnetic ground state of our exfoliated flake to a zigzag AFM order (Fig.~\ref{Figure3}g). Indeed, at low field, $G$ exhibits a small drop as $T$ decreases below $T_N$, which is most visible in thicker flakes (see Supporting Information, Figure S5). Such kinks are typically observed in MTJs with AFM interlayer order and attributed to a spin filter effect~\cite{Klein2018,Wang2018,Kim2019a,Klein2019}. This suggests that the ground state of $\alpha$-RuCl$_3$ flakes present both intra- and interlayer AFM order, in agreement with observations in the bulk \cite{Johnson2015,Banerjee2016,Cao2016}. At intermediate field, bulk crystals have been found to undergo a field-induced phase transition from this magnetically ordered state to a partially polarized quantum disordered state (QDS)~\cite{Johnson2015,Sears2017,Baek2017,Sahasrabudhe2020}. While the exact nature of this state is still under debate, it is typically characterized by a gapped spin-excitation continuum~\cite{Sahasrabudhe2020,Ponomaryov2020} and quantum fluctuations that prevent complete spin alignment. As a result, magnetic saturation can only be approached asymptotically with increasing $B$. This provides an explanation for the steadily increasing $G(B)$ we observe at high field (Fig.~\ref{Figure3}a), which contrasts with the magnetoconductance plateaus typically observed in other layered transition metal trihalides \cite{Wang2019,Klein2018}. In this high-field regime, the decrease of $G$ with increasing temperature (Fig.~\ref{Figure3}c) can be interpreted as a reduction of the spin polarization due to enhanced thermal fluctuations, ultimately resulting in a paramagnetic (PM) state. The minimum in $dG/dT$ (Fig.~\ref{Figure3}d) appears to demarcate the crossover between the PM and gapped QDS states. 

Next, we discuss the effect of the in-plane magnetic field orientation on the magnetic phase boundary. As can be seen by comparing Figures~\ref{Figure3}e and f, $B_c$ is slightly higher for $B||a$ (perpendicular to the Ru-Ru bonds) than for $B||b$ (parallel to the Ru-Ru bonds). This is consistent with our observation that $\delta G_b>\delta G_a$, which suggests that the magnetic susceptibility is higher for $B||b$. In contrast, in many bulk samples $B_c$ exhibits a 6-fold rotational symmetry and is maximum when the field is parallel to the Ru-Ru bonds~\cite{Tanaka2022,Balz2021}. However, Mi \emph{et al.} have found that this can be sample-dependent, having observed a maximum $B_c$ perpendicular to the Ru-Ru bonds for one of their bulk samples~\cite{Mi2021}. It is worth noting that several of these bulk samples exhibit one or several ordered phases below $B_c$. Some of our magnetoconductance measurements also indicate the presence of an intermediate phase when $B||a$. It is most visible in devices with thicker flakes, in particular $N = 7$ L (see Supporting Information, Figure  S5i). This phase might have the same origin as the narrow ZZ2~\cite{Balz2021} or X~\cite{Tanaka2022,Mi2021} phase observed in bulk samples (also most prominent for field perpendicular to the Ru-Ru bonds), in which the magnetic structure adopts a 6-layer stacking, instead of the 3-layer stacking at lower fields~\cite{Balz2021,Cen2022,Tanaka2022,Guo2016}.

In general, we note that the values of $T_N$ and $B_c$ we measure are larger than those typically reported for high-quality bulk samples~\cite{Banerjee2017,Sears2017,Kasahara2018,Tanaka2022,Balz2021} ($T_N \sim 7$ K, $B_c \sim 6 - 8$ T), but similar to those observed in samples with a high density of stacking fault~\cite{Sears2015,Kubota2015,Johnson2015,Cao2016,Mi2021}  ($T_N \sim 14$ K, $B_c \sim 8 - 10$ T). This enhanced $T_N$ has been linked to the two-layer stacking periodicity (ABAB) present in samples in powder form~\cite{Banerjee2016,Johnson2015} or mechanically deformed~\cite{Cao2016}. Indeed, owing to the weak van der Waals forces between individual layers, polytypes of $\alpha$-RuCl$_3$ have very small structural energy differences ($\textless$ 1 meV)~\cite{Kim2016}, making this material prone to stacking disorder. To determine the layer stacking order in our flakes, we performed high angle annular dark field scanning transmission electron microscopy (HAADF-STEM) on a $\sim 16$ nm-thick exfoliated flake at room temperature (see Supporting Information, Section II). While we predominantly observe STEM images consistent with the C2/m space group, we also observe several alternative structures which are not well-described by the C2/m structure or other known stacking orders of $\alpha$-RuCl$_3$. Instead, we posit that these regions contain disordered stacking and stacking faults as observed in other exfoliated 2D materials, such as MoTe$_2$~\cite{Hart2023} and TaS$_2$~\cite{Hovden2016}. These stacking faults may originate from extrinsic effects such as the strain applied on the flakes during their mechanical exfoliation or from intrinsic confinement effects on the stacking-dependent free energy~\cite{Hart2023}. The in-plane stacking domain size is $\textless$ 1 $\mu$m, and given that the area of our MTJ is typically on the order of 20 $\mu$m$^2$, we infer that they most certainly contain several stacking configurations. 

The presence of stacking faults can have an important impact on the low-temperature crystal structure. Bulk $\alpha$-RuCl$_3$ typically undergoes a crystallographic phase transition around 150 K from a monoclinic C2/m structure at room temperature to a rhombohedral R$\bar{3}$ structure at low temperature~\cite{Kubota2015,Mi2021,Reschke2017,Glamazda2017}. However, the two-fold rotational symmetry of our angle-dependent magnetoconductance measurements (Fig.~\ref{Figure2}b) indicates that the structure of our flakes remains in the C2/m phase at low temperature. Additionally, our temperature-dependent transport measurements (Fig.~\ref{Figure1}b, for example) show no sign of a structural transition, which should manifest as a jump in $G(T)$. A plausible explanation is that the stacking disorder effectively quenches the structural transition by pinning the crystal in the monoclinic phase. Similar effects have been observed in small $\alpha$-RuCl$_3$ samples~\cite{Cao2016} and other thin exfoliated flakes~\cite{Klein2019,Hart2023,He2018}. For $\alpha$-RuCl$_3$, variation in the stacking structures can affect not only the interlayer coupling, but also several of the intralayer interactions such as the Kitaev and $\Gamma$ interactions~\cite{Kim2016}. Hence, the larger values of $T_N$ and $B_c$ we observe likely stem from the frozen-in monoclinic phase of our exfoliated flakes.

\begin{figure*}[hbtp]
	\includegraphics[width=\textwidth]{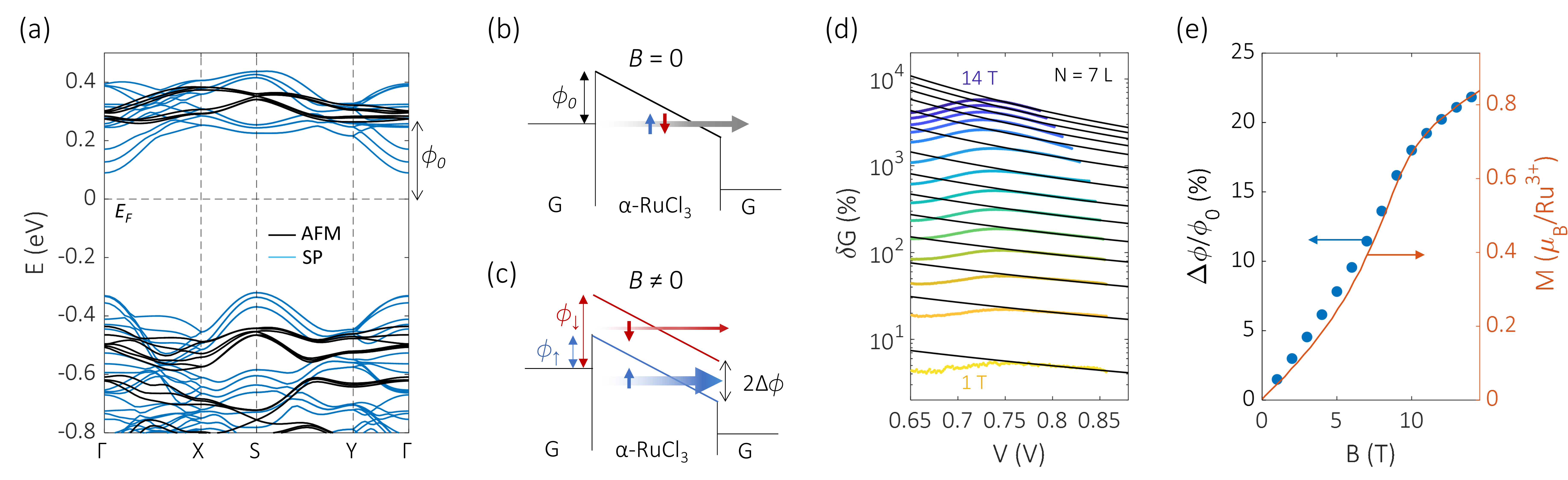}
	\caption{{\bf Electronic structure and origin of the magnetoconductance in $\alpha$-RuCl$_3$.} {\bf a} DFT calculations of the band structure of trilayer $\alpha$-RuCl$_3$ with C2/m stacking with zigzag AFM (black lines) or spin polarized (PS, blue lines) magnetic order. The energy barrier height $\phi_0$ between the bottom of the $\alpha$-RuCl$_3$ upper band and the Fermi energy $E_F$ is indicated. {\bf b,c} Energy band diagrams of graphite/$\alpha$-RuCl$_3$/graphite tunnel junctions ({\bf b}) at zero field and ({\bf c}) under in-plane magnetic field $B$. $\phi_{\uparrow,\downarrow}$ represents the energy barrier height experienced by majority- and minority-spin electrons. $2\Delta\phi=\phi_\downarrow-\phi_\uparrow$  is the total spin splitting energy between the two bands. The width of the arrows represents the magnitude of tunneling current. {\bf d} Magnetoconductance $\delta G$ as a function of voltage $V$ across a $\alpha$-RuCl$_3$ flake with $N = 7$ L at 1.8 K and selected values of $B$. The black lines are fits to the data using a spin-dependent FN tunneling model. Each fit yields a single fitting parameter, $\Delta\phi$. {\bf e} Left $y$ axis: $\Delta\phi / \phi_0$ as a function of $B$ (blue dots), where $\Delta\phi$ is obtained from the fits in {\bf d} and $\phi_0 = 0.26$ eV. Right $y$ axis: Magnetization $M$ as a function of in-plane magnetic field $B$ measured from a single crystal of $\alpha$-RuCl$_3$ with monoclinic C2/m structure (orange line, taken from Ref.~\cite{Johnson2015}). }
	\label{Figure4}
\end{figure*}

Finally, we use this knowledge to gain insight into the microscopic mechanism governing the magnetoconductance in our MTJs. For that purpose, we carried out density function theory (DFT) calculations of the band structure of trilayer $\alpha$-RuCl$_3$ with a C2/m stacking order (see Methods for calculations details). Initial calculations were performed with graphene layers on both sides of $\alpha$-RuCl$_3$, but they failed to capture the insulating behavior of $\alpha$-RuCl$_3$ found in our experiment (see Supporting Information, Section III). Since we are interested in understanding the effect of the external field on the $\alpha$-RuCl$_3$ bands, we computed the band structures of $\alpha$-RuCl$_3$ without graphene in the zigzag AFM and spin polarized (SP) magnetic configurations (Fig.~\ref{Figure4}a). In this case, we find that the Fermi energy ($E_F$) lies in the gap. The upper bands are composed of $J_\mathrm{eff} = 1/2$ while the lower bands are a combination of $J_\mathrm{eff} = 1/2$ and 3/2. However, since the precise location of $E_F$ cannot be determined numerically, we impose it to match the experimentally obtained value of $\phi_0$, i.e. the energy difference between the bottom of the $J_\mathrm{eff} = 1/2$ bands and $E_F$ at zero field. In the zero-field AFM configuration, $\alpha$-RuCl$_3$ bands are quite flat and without clear spin splitting, so electron tunneling is expected to be spin-independent (Fig.~\ref{Figure4}b). In the SP phase, the conduction band is completely spin-polarized, and majority-spin electrons experience a significantly reduced energy barrier. The relative energy shift of the potential barrier ($\Delta \phi/\phi_0$) varies with the momentum, ranging from  15\% around the X point and 65\% at the $\Gamma$ point. This magnetically dependent lowering of the energy barrier leads to an exponential increase of the tunneling current, thus explaining, at least qualitatively, the large magnetoconductance that we observe. 

We can quantitively compare our experimental results to these DFT calculations by analyzing our measurements with a simple spin-dependent FN tunneling model~\cite{Wang2021,Kim2018,Esaki1967}. In this phenomenological model, tunneling electrons with spin up and down experience tilted energy barriers with different heights $\phi_{\uparrow,\downarrow}$ (Fig.~\ref{Figure4}c, see Supporting Information, Section IV). We assume, as done by Wang \emph{et al.}~\cite{Wang2021}, that the field-induced spin splitting $\Delta\phi$ of the conduction band is symmetrical around the zero-field barrier height, i.e., $\phi_{\uparrow,\downarrow} =  \phi_0\pm \Delta \phi(B)$.  Figure~\ref{Figure4}d shows that this simple model captures well the decrease of the magnetoconductance at large bias for different values of $B$. From each curve, we extract a single fitting parameter $\Delta\phi(B)$, which we normalized by $\phi_0$ in Figure~\ref{Figure4}e. The maximum relative change extracted at $B = 14$ T is on the order of 20\%, which is comparable to our DFT predictions. Interestingly, the spin splitting energy $\Delta\phi(B)$ follows closely the magnetization curve $M(B)$ measured in bulk $\alpha$-RuCl$_3$ with monoclinic structure~\cite{Johnson2015}. This suggests, as previously reported for CrBr$_3$~\cite{Wang2021},  that the spin splitting energy is linearly proportional to the magnetization, which elucidates the relation between the tunneling magnetoconductance of $\alpha$-RuCl$_3$ and its magnetization.

In summary, our study of $\alpha$-RuCl$_3$ demonstrates that angle-dependent TMR measurements can  provide a multitude of information on the magnetic, electronic and crystal properties of ultrathin frustrated magnets. We find that they exhibit a strong easy-plane magnetic anisotropy with a two-fold in-plane symmetry, indicating that their structure remains in the monoclinic phase at low temperature. As a result, exfoliated flakes present a zigzag AFM magnetic ground state with enhanced critical values ($T_N$, $B_c$) compared bulk $\alpha$-RuCl$_3$ with rhombohedral stacking. These results demonstrate the influence of stacking order on the magnetic properties of van der Waals materials. This opens up the possibility of generating exotic magnetic phases, such as QSLs, by controlling the layer stacking via, for instance, hydrostatic pressure~\cite{Li2019,Song2019} or twisting~\cite{Song2021}. As such, our work paves the way for the development of spintronic devices exploiting emergent excitations in these unconventional phases.

\section{Methods}

{\bf Crystal synthesis.} Single-crystal RuCl$_3$ was synthesized from $\alpha$-RuCl$_3$ powder provided by Furuya Metals (Japan). The powder was sealed in a quartz ampoule that had been purged with argon and then placed under vacuum. The ampoule was heated to 1060ºC at 1.6°C/min. It was held at 1060°C for 12 hours before being slowly cooled to 600°C at 6 °C/hr. Crystals grew via chemical vapor transport as shiny black plates. These were characterized via x-ray diffraction and magnetic susceptibility to determine phase purity and sample quality. A single peak in the susceptibility was observed around 7~K. This peak, along with the absence of an additional peak around 14~K, has been shown to indicate low stacking fault density~\cite{Cao2016}. 

{\bf Device fabrication and transport measurements.} $\alpha$-RuCl$_3$ flakes were mechanically exfoliated from the bulk crystal. Tunnel junctions of hBN/graphite/$\alpha$-RuCl$_3$/graphite were assembled using a dry pick-up technique using stamps of PDMS/PC. We note that the relative angle between the crystalline axes of the flakes is not controlled. Some of the heterostructures were fabricated inside a glove box filled with $N_2$, others in air. No difference was observed in the quality of the junction, which indicates that $\alpha$-RuCl$_3$ is air stable. The heterostructures were deposited onto a silicon substrate with a 285 nm oxide layer and prepatterned Ti/Au leads to contact the graphite electrodes. Transport measurements were performed in a cryostat from Quantum Design (Dynacool Physical Properties Measurement System) equipped with a sample rotator. We used a combination of DC voltage source (Yokogawa GS200), multimeter (Agilent 34410A) and current preamplifier (Ithaco 1211) to measure the conductance of the junction, as well as the graphite flakes. To ensure an accurate measurement of the tunnel junction and avoid contribution from its graphite contacts, we verified that the junction resistance was always much greater than that of the graphite flakes.

{\bf DFT calculations.} The density functional theory (DFT) calculations are performed with Vienna \emph{ab initio} Simulation Packages (VASP)~\cite{Kresse1993} with the projector augmented wave potential~\cite{Blochl1994} and the Perdew-Burke-Ernzerhof (PBE)~\cite{Perdew1996} exchange-correlation functionals. Three layers of $\alpha$-RuCl$_3$ with C2/m type stacking type (rectangular unit cell) are considered.  A vacuum layer of 15 \AA~is included to avoid interactions between images due to the periodic boundary conditions. The energy cutoff for the plane-wave basis is 400 eV and the  $k$-point mesh is $6\times 3\times 1$.  Both the spin-orbit coupling (SOC) effect and the onsite effective Coulomb interactions $U = 1.5$ eV are included~\cite{Kim2015}. For the zigzag AFM calculations, the magnetic structure has intralayer AFM zigzag chains and is AFM between layers. The magnetic moments are in the $ac$ plane with a small $c$-axis component.  For spin-polarized (SP) calculations, the magnetic moments are along the $a$ axis.

\section{Acknowledgments}

X.L and H.Y.K acknowledge the support of the Natural Sciences and Engineering Research Council of Canada  (Discovery Grants No. RGPIN2022-04601). H.Y. K. also acknowledges the support of the Canadian Institute for Advanced Research and the Canada Research Chairs Program. J. L. H. and J. J. C. acknowledge generous support from Gordon and Betty Moore Foundation under the EPiQS program, GBMF9062.01. The microscopy facilities at Cornell are supported through the NSF MRSEC program (DMR-1719875). S..E.N. and J.Q.Y. was supported by the U.S. Department of Energy, Office of Science, National Quantum Information Science Research Centers, Quantum Science Center, Office of Science, Basic Energy Sciences, Materials Science and Engineering Division; this support contributed to the growth and characterization of the crystals used in this work. D.G.M. acknowledges support from the Gordon and Betty Moore Foundation’s EPiQS Initiative, Grant GBMF9069.  M.M., J.A.Q. and B.R. acknowledge research funding from the Canada First Research Excellence Fund and the Natural Sciences and Engineering Research Council of Canada. B.R. also acknowledges support from the Canada Research Chair program, the Canada Foundation for Innovation and the FRQNT. We thank Gabriel Laliberté, Christian Lupien and Jordan Baglo for technical support, and acknowledge helpful discussions with Quentin Barthélemy. 

\begin{suppinfo}
Supporting Information: Additional measurements on $\alpha$-RuCl$_3$ tunnel junctions, high angle annular dark field scanning transmission electron microscopy, additional DFT calculations, spin-dependent Fowler-Nordheim tunneling model (PDF).

\end{suppinfo}

\bibliography{RUCl3_ref}

\providecommand{\latin}[1]{#1}
\makeatletter
\providecommand{\doi}
  {\begingroup\let\do\@makeother\dospecials
  \catcode`\{=1 \catcode`\}=2 \doi@aux}
\providecommand{\doi@aux}[1]{\endgroup\texttt{#1}}
\makeatother
\providecommand*\mcitethebibliography{\thebibliography}
\csname @ifundefined\endcsname{endmcitethebibliography}  {\let\endmcitethebibliography\endthebibliography}{}
\begin{mcitethebibliography}{80}
\providecommand*\natexlab[1]{#1}
\providecommand*\mciteSetBstSublistMode[1]{}
\providecommand*\mciteSetBstMaxWidthForm[2]{}
\providecommand*\mciteBstWouldAddEndPuncttrue
  {\def\EndOfBibitem{\unskip.}}
\providecommand*\mciteBstWouldAddEndPunctfalse
  {\let\EndOfBibitem\relax}
\providecommand*\mciteSetBstMidEndSepPunct[3]{}
\providecommand*\mciteSetBstSublistLabelBeginEnd[3]{}
\providecommand*\EndOfBibitem{}
\mciteSetBstSublistMode{f}
\mciteSetBstMaxWidthForm{subitem}{(\alph{mcitesubitemcount})}
\mciteSetBstSublistLabelBeginEnd
  {\mcitemaxwidthsubitemform\space}
  {\relax}
  {\relax}

\bibitem[Savary and Balents(2017)Savary, and Balents]{Savary2017}
Savary,~L.; Balents,~L. {Quantum spin liquids: a review}. \emph{Reports on Progress in Physics} \textbf{2017}, \emph{80}, 016502\relax
\mciteBstWouldAddEndPuncttrue
\mciteSetBstMidEndSepPunct{\mcitedefaultmidpunct}
{\mcitedefaultendpunct}{\mcitedefaultseppunct}\relax
\EndOfBibitem
\bibitem[Anderson(1973)]{Anderson1973}
Anderson,~P.~W. {Resonating valence bonds: A new kind of insulator?} \emph{Materials Research Bulletin} \textbf{1973}, \emph{8}, 153--160\relax
\mciteBstWouldAddEndPuncttrue
\mciteSetBstMidEndSepPunct{\mcitedefaultmidpunct}
{\mcitedefaultendpunct}{\mcitedefaultseppunct}\relax
\EndOfBibitem
\bibitem[Balents(2010)]{Balents2010}
Balents,~L. {Spin liquids in frustrated magnets}. \emph{Nature} \textbf{2010}, \emph{464}, 199--208\relax
\mciteBstWouldAddEndPuncttrue
\mciteSetBstMidEndSepPunct{\mcitedefaultmidpunct}
{\mcitedefaultendpunct}{\mcitedefaultseppunct}\relax
\EndOfBibitem
\bibitem[Kitaev(2006)]{Kitaev2006}
Kitaev,~A. {Anyons in an exactly solved model and beyond}. \emph{Annals of Physics} \textbf{2006}, \emph{321}, 2--111\relax
\mciteBstWouldAddEndPuncttrue
\mciteSetBstMidEndSepPunct{\mcitedefaultmidpunct}
{\mcitedefaultendpunct}{\mcitedefaultseppunct}\relax
\EndOfBibitem
\bibitem[Nayak \latin{et~al.}(2008)Nayak, Simon, Stern, Freedman, and {Das Sarma}]{Nayak2008}
Nayak,~C.; Simon,~S.~H.; Stern,~A.; Freedman,~M.; {Das Sarma},~S. {Non-Abelian anyons and topological quantum computation}. \emph{Reviews of Modern Physics} \textbf{2008}, \emph{80}, 1083--1159\relax
\mciteBstWouldAddEndPuncttrue
\mciteSetBstMidEndSepPunct{\mcitedefaultmidpunct}
{\mcitedefaultendpunct}{\mcitedefaultseppunct}\relax
\EndOfBibitem
\bibitem[Jackeli and Khaliullin(2009)Jackeli, and Khaliullin]{Jackeli2009}
Jackeli,~G.; Khaliullin,~G. {Mott Insulators in the Strong Spin-Orbit Coupling Limit: From Heisenberg to a Quantum Compass and Kitaev Models}. \emph{Physical Review Letters} \textbf{2009}, \emph{102}, 017205\relax
\mciteBstWouldAddEndPuncttrue
\mciteSetBstMidEndSepPunct{\mcitedefaultmidpunct}
{\mcitedefaultendpunct}{\mcitedefaultseppunct}\relax
\EndOfBibitem
\bibitem[Takagi \latin{et~al.}(2019)Takagi, Takayama, Jackeli, Khaliullin, and Nagler]{Takagi2019}
Takagi,~H.; Takayama,~T.; Jackeli,~G.; Khaliullin,~G.; Nagler,~S.~E. {Concept and realization of Kitaev quantum spin liquids}. \emph{Nature Reviews Physics} \textbf{2019}, \emph{1}, 264--280\relax
\mciteBstWouldAddEndPuncttrue
\mciteSetBstMidEndSepPunct{\mcitedefaultmidpunct}
{\mcitedefaultendpunct}{\mcitedefaultseppunct}\relax
\EndOfBibitem
\bibitem[Plumb \latin{et~al.}(2014)Plumb, Clancy, Sandilands, Shankar, Hu, Burch, Kee, and Kim]{Plumb2014}
Plumb,~K.~W.; Clancy,~J.~P.; Sandilands,~L.~J.; Shankar,~V.~V.; Hu,~Y.~F.; Burch,~K.~S.; Kee,~H.-Y.; Kim,~Y.-J. {$\alpha$-RuCl$_3$: A spin-orbit assisted Mott insulator on a honeycomb lattice}. \emph{Physical Review B} \textbf{2014}, \emph{90}, 041112\relax
\mciteBstWouldAddEndPuncttrue
\mciteSetBstMidEndSepPunct{\mcitedefaultmidpunct}
{\mcitedefaultendpunct}{\mcitedefaultseppunct}\relax
\EndOfBibitem
\bibitem[Johnson \latin{et~al.}(2015)Johnson, Williams, Haghighirad, Singleton, Zapf, Manuel, Mazin, Li, Jeschke, Valent{\'{i}}, and Coldea]{Johnson2015}
Johnson,~R.~D.; Williams,~S.~C.; Haghighirad,~A.~A.; Singleton,~J.; Zapf,~V.; Manuel,~P.; Mazin,~I.~I.; Li,~Y.; Jeschke,~H.~O.; Valent{\'{i}},~R.; Coldea,~R. {Monoclinic crystal structure of $\alpha$-RuCl$_3$ and the zigzag antiferromagnetic ground state}. \emph{Physical Review B} \textbf{2015}, \emph{92}, 235119\relax
\mciteBstWouldAddEndPuncttrue
\mciteSetBstMidEndSepPunct{\mcitedefaultmidpunct}
{\mcitedefaultendpunct}{\mcitedefaultseppunct}\relax
\EndOfBibitem
\bibitem[Sears \latin{et~al.}(2015)Sears, Songvilay, Plumb, Clancy, Qiu, Zhao, Parshall, and Kim]{Sears2015}
Sears,~J.~A.; Songvilay,~M.; Plumb,~K.~W.; Clancy,~J.~P.; Qiu,~Y.; Zhao,~Y.; Parshall,~D.; Kim,~Y.~J. {Magnetic order in $\alpha$-RuCl$_3$: A honeycomb-lattice quantum magnet with strong spin-orbit coupling}. \emph{Physical Review B} \textbf{2015}, \emph{91}, 144420\relax
\mciteBstWouldAddEndPuncttrue
\mciteSetBstMidEndSepPunct{\mcitedefaultmidpunct}
{\mcitedefaultendpunct}{\mcitedefaultseppunct}\relax
\EndOfBibitem
\bibitem[Majumder \latin{et~al.}(2015)Majumder, Schmidt, Rosner, Tsirlin, Yasuoka, and Baenitz]{Majumder2015}
Majumder,~M.; Schmidt,~M.; Rosner,~H.; Tsirlin,~A.~A.; Yasuoka,~H.; Baenitz,~M. {Anisotropic Ru3+ 4d5 magnetisminthe $\alpha$-RuCl$_3$ honeycomb system: Susceptibility, specific heat, and zero-field NMR}. \emph{Physical Review B} \textbf{2015}, \emph{91}, 180401\relax
\mciteBstWouldAddEndPuncttrue
\mciteSetBstMidEndSepPunct{\mcitedefaultmidpunct}
{\mcitedefaultendpunct}{\mcitedefaultseppunct}\relax
\EndOfBibitem
\bibitem[Banerjee \latin{et~al.}(2017)Banerjee, Yan, Knolle, Bridges, Stone, Lumsden, Mandrus, Tennant, Moessner, and Nagler]{Banerjee2017}
Banerjee,~A.; Yan,~J.; Knolle,~J.; Bridges,~C.~A.; Stone,~M.~B.; Lumsden,~M.~D.; Mandrus,~D.~G.; Tennant,~D.~A.; Moessner,~R.; Nagler,~S.~E. {Neutron scattering in the proximate quantum spin liquid $\alpha$-RuCl$_3$}. \emph{Science} \textbf{2017}, \emph{356}, 1055--1059\relax
\mciteBstWouldAddEndPuncttrue
\mciteSetBstMidEndSepPunct{\mcitedefaultmidpunct}
{\mcitedefaultendpunct}{\mcitedefaultseppunct}\relax
\EndOfBibitem
\bibitem[Banerjee \latin{et~al.}(2016)Banerjee, Bridges, Yan, Aczel, Li, Stone, Granroth, Lumsden, Yiu, Knolle, Bhattacharjee, Kovrizhin, Moessner, Tennant, Mandrus, and Nagler]{Banerjee2016}
Banerjee,~A. \latin{et~al.}  {Proximate Kitaev quantum spin liquid behaviour in a honeycomb magnet}. \emph{Nature Materials} \textbf{2016}, \emph{15}, 733--740\relax
\mciteBstWouldAddEndPuncttrue
\mciteSetBstMidEndSepPunct{\mcitedefaultmidpunct}
{\mcitedefaultendpunct}{\mcitedefaultseppunct}\relax
\EndOfBibitem
\bibitem[Nasu \latin{et~al.}(2016)Nasu, Knolle, Kovrizhin, Motome, and Moessner]{Nasu2016}
Nasu,~J.; Knolle,~J.; Kovrizhin,~D.~L.; Motome,~Y.; Moessner,~R. {Fermionic response from fractionalization in an insulating two-dimensional magnet}. \emph{Nature Physics} \textbf{2016}, \emph{12}, 912--915\relax
\mciteBstWouldAddEndPuncttrue
\mciteSetBstMidEndSepPunct{\mcitedefaultmidpunct}
{\mcitedefaultendpunct}{\mcitedefaultseppunct}\relax
\EndOfBibitem
\bibitem[Zheng \latin{et~al.}(2017)Zheng, Ran, Li, Wang, Wang, Liu, Liu, Normand, Wen, and Yu]{Zheng2017}
Zheng,~J.; Ran,~K.; Li,~T.; Wang,~J.; Wang,~P.; Liu,~B.; Liu,~Z.-X.; Normand,~B.; Wen,~J.; Yu,~W. {Gapless Spin Excitations in the Field-Induced Quantum Spin Liquid Phase of $\alpha$-RuCl$_3$}. \emph{Physical Review Letters} \textbf{2017}, \emph{119}, 227208\relax
\mciteBstWouldAddEndPuncttrue
\mciteSetBstMidEndSepPunct{\mcitedefaultmidpunct}
{\mcitedefaultendpunct}{\mcitedefaultseppunct}\relax
\EndOfBibitem
\bibitem[Sears \latin{et~al.}(2017)Sears, Zhao, Xu, Lynn, and Kim]{Sears2017}
Sears,~J.~A.; Zhao,~Y.; Xu,~Z.; Lynn,~J.~W.; Kim,~Y.~J. {Phase diagram of $\alpha$-RuCl$_3$ in an in-plane magnetic field}. \emph{Physical Review B} \textbf{2017}, \emph{95}, 180411\relax
\mciteBstWouldAddEndPuncttrue
\mciteSetBstMidEndSepPunct{\mcitedefaultmidpunct}
{\mcitedefaultendpunct}{\mcitedefaultseppunct}\relax
\EndOfBibitem
\bibitem[Wolter \latin{et~al.}(2017)Wolter, Corredor, Janssen, Nenkov, Sch{\"{o}}necker, Do, Choi, Albrecht, Hunger, Doert, Vojta, and B{\"{u}}chner]{Wolter2017}
Wolter,~A.~U.; Corredor,~L.~T.; Janssen,~L.; Nenkov,~K.; Sch{\"{o}}necker,~S.; Do,~S.~H.; Choi,~K.~Y.; Albrecht,~R.; Hunger,~J.; Doert,~T.; Vojta,~M.; B{\"{u}}chner,~B. {Field-induced quantum criticality in the Kitaev system $\alpha$-RuCl$_3$}. \emph{Physical Review B} \textbf{2017}, \emph{96}, 41405\relax
\mciteBstWouldAddEndPuncttrue
\mciteSetBstMidEndSepPunct{\mcitedefaultmidpunct}
{\mcitedefaultendpunct}{\mcitedefaultseppunct}\relax
\EndOfBibitem
\bibitem[Wang \latin{et~al.}(2017)Wang, Reschke, H{\"{u}}vonen, Do, Choi, Gensch, Nagel, R{\~{o}}{\~{o}}m, and Loidl]{Wang2017}
Wang,~Z.; Reschke,~S.; H{\"{u}}vonen,~D.; Do,~S.~H.; Choi,~K.~Y.; Gensch,~M.; Nagel,~U.; R{\~{o}}{\~{o}}m,~T.; Loidl,~A. {Magnetic Excitations and Continuum of a Possibly Field-Induced Quantum Spin Liquid in $\alpha$-RuCl$_3$}. \emph{Physical Review Letters} \textbf{2017}, \emph{119}\relax
\mciteBstWouldAddEndPuncttrue
\mciteSetBstMidEndSepPunct{\mcitedefaultmidpunct}
{\mcitedefaultendpunct}{\mcitedefaultseppunct}\relax
\EndOfBibitem
\bibitem[Baek \latin{et~al.}(2017)Baek, Do, Choi, Kwon, Wolter, Nishimoto, {Van Den Brink}, and B{\"{o}}chner]{Baek2017}
Baek,~S.~H.; Do,~S.~H.; Choi,~K.~Y.; Kwon,~Y.~S.; Wolter,~A.~U.; Nishimoto,~S.; {Van Den Brink},~J.; B{\"{o}}chner,~B. {Evidence for a field-induced quantum spin liquid in $\alpha$-RuCl$_3$}. \emph{Physical Review Letters} \textbf{2017}, \emph{119}\relax
\mciteBstWouldAddEndPuncttrue
\mciteSetBstMidEndSepPunct{\mcitedefaultmidpunct}
{\mcitedefaultendpunct}{\mcitedefaultseppunct}\relax
\EndOfBibitem
\bibitem[Kasahara \latin{et~al.}(2018)Kasahara, Ohnishi, Mizukami, Tanaka, Ma, Sugii, Kurita, Tanaka, Nasu, Motome, Shibauchi, and Matsuda]{Kasahara2018}
Kasahara,~Y.; Ohnishi,~T.; Mizukami,~Y.; Tanaka,~O.; Ma,~S.; Sugii,~K.; Kurita,~N.; Tanaka,~H.; Nasu,~J.; Motome,~Y.; Shibauchi,~T.; Matsuda,~Y. {Majorana quantization and half-integer thermal quantum Hall effect in a Kitaev spin liquid}. \emph{Nature} \textbf{2018}, \emph{559}, 227--231\relax
\mciteBstWouldAddEndPuncttrue
\mciteSetBstMidEndSepPunct{\mcitedefaultmidpunct}
{\mcitedefaultendpunct}{\mcitedefaultseppunct}\relax
\EndOfBibitem
\bibitem[Kim \latin{et~al.}(2022)Kim, Yuan, and Kim]{Kim2022}
Kim,~S.; Yuan,~B.; Kim,~Y.~J. {$\alpha$-RuCl$_3$ and other Kitaev materials}. \emph{APL Materials} \textbf{2022}, \emph{10}, 080903\relax
\mciteBstWouldAddEndPuncttrue
\mciteSetBstMidEndSepPunct{\mcitedefaultmidpunct}
{\mcitedefaultendpunct}{\mcitedefaultseppunct}\relax
\EndOfBibitem
\bibitem[Yamashita \latin{et~al.}(2020)Yamashita, Gouchi, Uwatoko, Kurita, and Tanaka]{Yamashita2020}
Yamashita,~M.; Gouchi,~J.; Uwatoko,~Y.; Kurita,~N.; Tanaka,~H. {Sample dependence of half-integer quantized thermal Hall effect in the Kitaev spin-liquid candidate $\alpha$-RuCl$_3$}. \emph{Physical Review B} \textbf{2020}, \emph{102}, 220404\relax
\mciteBstWouldAddEndPuncttrue
\mciteSetBstMidEndSepPunct{\mcitedefaultmidpunct}
{\mcitedefaultendpunct}{\mcitedefaultseppunct}\relax
\EndOfBibitem
\bibitem[Lefran{\c{c}}ois \latin{et~al.}(2022)Lefran{\c{c}}ois, Grissonnanche, Baglo, Lampen-Kelley, Yan, Balz, Mandrus, Nagler, Kim, Kim, Doiron-Leyraud, and Taillefer]{Lefrancois2022}
Lefran{\c{c}}ois,~E.; Grissonnanche,~G.; Baglo,~J.; Lampen-Kelley,~P.; Yan,~J.~Q.; Balz,~C.; Mandrus,~D.; Nagler,~S.~E.; Kim,~S.; Kim,~Y.~J.; Doiron-Leyraud,~N.; Taillefer,~L. {Evidence of a Phonon Hall Effect in the Kitaev Spin Liquid Candidate $\alpha$-RuCl$_3$}. \emph{Physical Review X} \textbf{2022}, \emph{12}, 021025\relax
\mciteBstWouldAddEndPuncttrue
\mciteSetBstMidEndSepPunct{\mcitedefaultmidpunct}
{\mcitedefaultendpunct}{\mcitedefaultseppunct}\relax
\EndOfBibitem
\bibitem[Du \latin{et~al.}(2019)Du, Huang, Wang, Wang, Yang, Tang, Liao, Shi, Shi, Zhou, Zhang, and Zhang]{Du2019}
Du,~L.; Huang,~Y.; Wang,~Y.; Wang,~Q.; Yang,~R.; Tang,~J.; Liao,~M.; Shi,~D.; Shi,~Y.; Zhou,~X.; Zhang,~Q.; Zhang,~G. {2D proximate quantum spin liquid state in atomic-thin $\alpha$-RuCl$_3$}. \emph{2D Materials} \textbf{2019}, \emph{6}, 015014\relax
\mciteBstWouldAddEndPuncttrue
\mciteSetBstMidEndSepPunct{\mcitedefaultmidpunct}
{\mcitedefaultendpunct}{\mcitedefaultseppunct}\relax
\EndOfBibitem
\bibitem[Lin \latin{et~al.}(2020)Lin, Ran, Zheng, Xu, Gao, Wen, Yu, Li, and Xi]{Lin2020}
Lin,~D.; Ran,~K.; Zheng,~H.; Xu,~J.; Gao,~L.; Wen,~J.; Yu,~S.~L.; Li,~J.~X.; Xi,~X. {Anisotropic scattering continuum induced by crystal symmetry reduction in atomically thin $\alpha$-RuCl$_3$}. \emph{Physical Review B} \textbf{2020}, \emph{101}, 45419\relax
\mciteBstWouldAddEndPuncttrue
\mciteSetBstMidEndSepPunct{\mcitedefaultmidpunct}
{\mcitedefaultendpunct}{\mcitedefaultseppunct}\relax
\EndOfBibitem
\bibitem[Zhou \latin{et~al.}(2019)Zhou, Wang, Osterhoudt, Lampen-Kelley, Mandrus, He, Burch, and Henriksen]{Zhou2019}
Zhou,~B.; Wang,~Y.; Osterhoudt,~G.~B.; Lampen-Kelley,~P.; Mandrus,~D.; He,~R.; Burch,~K.~S.; Henriksen,~E.~A. {Possible structural transformation and enhanced magnetic fluctuations in exfoliated $\alpha$-RuCl$_3$}. \emph{Journal of Physics and Chemistry of Solids} \textbf{2019}, \emph{128}, 291--295\relax
\mciteBstWouldAddEndPuncttrue
\mciteSetBstMidEndSepPunct{\mcitedefaultmidpunct}
{\mcitedefaultendpunct}{\mcitedefaultseppunct}\relax
\EndOfBibitem
\bibitem[Dai \latin{et~al.}(2020)Dai, Yu, Zhou, {A Tenney}, Lampen-Kelley, Yan, Mandrus, {A Henriksen}, Zang, Pohl, and Sadowski]{Dai2020}
Dai,~Z.; Yu,~J.~X.; Zhou,~B.; {A Tenney},~S.; Lampen-Kelley,~P.; Yan,~J.; Mandrus,~D.; {A Henriksen},~E.; Zang,~J.; Pohl,~K.; Sadowski,~J.~T. {Crystal structure reconstruction in the surface monolayer of the quantum spin liquid candidate $\alpha$-RuCl$_3$}. \emph{2D Materials} \textbf{2020}, \emph{7}, 035004\relax
\mciteBstWouldAddEndPuncttrue
\mciteSetBstMidEndSepPunct{\mcitedefaultmidpunct}
{\mcitedefaultendpunct}{\mcitedefaultseppunct}\relax
\EndOfBibitem
\bibitem[Vatansever \latin{et~al.}(2019)Vatansever, Sarikurt, Ersan, Kadioglu, {{\"{U}}zengi Akt{\"{u}}rk}, Y{\"{u}}ksel, Ataca, Akt{\"{u}}rk, and Aklncl]{Vatansever2019}
Vatansever,~E.; Sarikurt,~S.; Ersan,~F.; Kadioglu,~Y.; {{\"{U}}zengi Akt{\"{u}}rk},~O.; Y{\"{u}}ksel,~Y.; Ataca,~C.; Akt{\"{u}}rk,~E.; Aklncl,~{\"{U}}. {Strain effects on electronic and magnetic properties of the monolayer $\alpha$-RuCl$_3$: A first-principles and Monte Carlo study}. \emph{Journal of Applied Physics} \textbf{2019}, \emph{125}, 083903\relax
\mciteBstWouldAddEndPuncttrue
\mciteSetBstMidEndSepPunct{\mcitedefaultmidpunct}
{\mcitedefaultendpunct}{\mcitedefaultseppunct}\relax
\EndOfBibitem
\bibitem[Zheng \latin{et~al.}(2023)Zheng, Jia, Ren, Yang, Wu, Shi, Tanigaki, and Du]{Zheng2023}
Zheng,~X.; Jia,~K.; Ren,~J.; Yang,~C.; Wu,~X.; Shi,~Y.; Tanigaki,~K.; Du,~R.~R. {Tunneling spectroscopic signatures of charge doping and associated Mott transition in $\alpha$-RuCl$_3$ in proximity to graphite}. \emph{Physical Review B} \textbf{2023}, \emph{107}, 195107\relax
\mciteBstWouldAddEndPuncttrue
\mciteSetBstMidEndSepPunct{\mcitedefaultmidpunct}
{\mcitedefaultendpunct}{\mcitedefaultseppunct}\relax
\EndOfBibitem
\bibitem[Yang \latin{et~al.}(2023)Yang, Goh, Sung, Ye, Biswas, Kaib, Dhakal, Yan, Li, Jiang, Chen, Lei, He, Valent{\'{i}}, Winter, Hovden, and Tsen]{Yang2023a}
Yang,~B. \latin{et~al.}  {Magnetic anisotropy reversal driven by structural symmetry-breaking in monolayer $\alpha$-RuCl$_3$}. \emph{Nature Materials} \textbf{2023}, \emph{22}, 50--57\relax
\mciteBstWouldAddEndPuncttrue
\mciteSetBstMidEndSepPunct{\mcitedefaultmidpunct}
{\mcitedefaultendpunct}{\mcitedefaultseppunct}\relax
\EndOfBibitem
\bibitem[Wang \latin{et~al.}(2020)Wang, Balgley, Gerber, Gray, Kumar, Lu, Yan, Fereidouni, Basnet, Yun, Suri, Kitadai, Taniguchi, Watanabe, Ling, Moodera, Lee, Churchill, Hu, Yang, Kim, Mandrus, Henriksen, and Burch]{Wang2020}
Wang,~Y. \latin{et~al.}  {Modulation Doping via a Two-Dimensional Atomic Crystalline Acceptor}. \emph{Nano Letters} \textbf{2020}, \emph{20}, 8446--8452\relax
\mciteBstWouldAddEndPuncttrue
\mciteSetBstMidEndSepPunct{\mcitedefaultmidpunct}
{\mcitedefaultendpunct}{\mcitedefaultseppunct}\relax
\EndOfBibitem
\bibitem[Mashhadi \latin{et~al.}(2019)Mashhadi, Kim, Kim, Weber, Taniguchi, Watanabe, Park, Lotsch, Smet, Burghard, and Kern]{Mashhadi2019}
Mashhadi,~S.; Kim,~Y.; Kim,~J.; Weber,~D.; Taniguchi,~T.; Watanabe,~K.; Park,~N.; Lotsch,~B.; Smet,~J.~H.; Burghard,~M.; Kern,~K. {Spin-Split Band Hybridization in Graphene Proximitized with $\alpha$-RuCl$_3$ Nanosheets}. \emph{Nano Letters} \textbf{2019}, \emph{19}, 4659--4665\relax
\mciteBstWouldAddEndPuncttrue
\mciteSetBstMidEndSepPunct{\mcitedefaultmidpunct}
{\mcitedefaultendpunct}{\mcitedefaultseppunct}\relax
\EndOfBibitem
\bibitem[Rizzo \latin{et~al.}(2020)Rizzo, Jessen, Sun, Ruta, Zhang, Yan, Xian, McLeod, Berkowitz, Watanabe, Taniguchi, Nagler, Mandrus, Rubio, Fogler, Millis, Hone, Dean, and Basov]{Rizzo2020}
Rizzo,~D.~J. \latin{et~al.}  {Charge-Transfer Plasmon Polaritons at Graphene/$\alpha$-RuCl$_3$ Interfaces}. \emph{Nano Letters} \textbf{2020}, \emph{20}, 8438--8445\relax
\mciteBstWouldAddEndPuncttrue
\mciteSetBstMidEndSepPunct{\mcitedefaultmidpunct}
{\mcitedefaultendpunct}{\mcitedefaultseppunct}\relax
\EndOfBibitem
\bibitem[Zhou \latin{et~al.}(2019)Zhou, Balgley, Lampen-Kelley, Yan, Mandrus, and Henriksen]{Zhou2019a}
Zhou,~B.; Balgley,~J.; Lampen-Kelley,~P.; Yan,~J.-Q.; Mandrus,~D.~G.; Henriksen,~E.~A. {Evidence for charge transfer and proximate magnetism in graphene–$\alpha$-RuCl$_3$ heterostructures}. \emph{Physical Review B} \textbf{2019}, \emph{100}, 165426\relax
\mciteBstWouldAddEndPuncttrue
\mciteSetBstMidEndSepPunct{\mcitedefaultmidpunct}
{\mcitedefaultendpunct}{\mcitedefaultseppunct}\relax
\EndOfBibitem
\bibitem[Souza \latin{et~al.}(2022)Souza, Deus, Brito, and Miwa]{Souza2022}
Souza,~P.~H.; Deus,~D. P.~A.; Brito,~W.~H.; Miwa,~R.~H. {Magnetic anisotropy energies and metal-insulator transitions in monolayers of $\alpha$-RuCl$_3$ and OsCl$_3$ on graphene}. \emph{Physical Review B} \textbf{2022}, \emph{106}, 155118\relax
\mciteBstWouldAddEndPuncttrue
\mciteSetBstMidEndSepPunct{\mcitedefaultmidpunct}
{\mcitedefaultendpunct}{\mcitedefaultseppunct}\relax
\EndOfBibitem
\bibitem[Biswas \latin{et~al.}(2019)Biswas, Li, Winter, Knolle, and Valent{\'{i}}]{Biswas2019}
Biswas,~S.; Li,~Y.; Winter,~S.~M.; Knolle,~J.; Valent{\'{i}},~R. {Electronic Properties of $\alpha$-RuCl$_3$ in Proximity to Graphene}. \emph{Physical Review Letters} \textbf{2019}, \emph{123}\relax
\mciteBstWouldAddEndPuncttrue
\mciteSetBstMidEndSepPunct{\mcitedefaultmidpunct}
{\mcitedefaultendpunct}{\mcitedefaultseppunct}\relax
\EndOfBibitem
\bibitem[Gerber \latin{et~al.}(2020)Gerber, Yao, Arias, and Kim]{Gerber2020}
Gerber,~E.; Yao,~Y.; Arias,~T.~A.; Kim,~E.~A. {Ab Initio Mismatched Interface Theory of Graphene on $\alpha$-RuCl$_3$: Doping and Magnetism}. \emph{Physical Review Letters} \textbf{2020}, \emph{124}, 106804\relax
\mciteBstWouldAddEndPuncttrue
\mciteSetBstMidEndSepPunct{\mcitedefaultmidpunct}
{\mcitedefaultendpunct}{\mcitedefaultseppunct}\relax
\EndOfBibitem
\bibitem[Wang \latin{et~al.}(2019)Wang, Gibertini, Dumcenco, Taniguchi, Watanabe, Giannini, and Morpurgo]{Wang2019}
Wang,~Z.; Gibertini,~M.; Dumcenco,~D.; Taniguchi,~T.; Watanabe,~K.; Giannini,~E.; Morpurgo,~A.~F. {Determining the phase diagram of atomically thin layered antiferromagnet CrCl$_3$}. \emph{Nature Nanotechnology} \textbf{2019}, \emph{14}, 1116--1122\relax
\mciteBstWouldAddEndPuncttrue
\mciteSetBstMidEndSepPunct{\mcitedefaultmidpunct}
{\mcitedefaultendpunct}{\mcitedefaultseppunct}\relax
\EndOfBibitem
\bibitem[Klein \latin{et~al.}(2018)Klein, MacNeill, Lado, Soriano, Navarro-Moratalla, Watanabe, Taniguchi, Manni, Canfield, Fern{\'{a}}ndez-Rossier, and Jarillo-Herrero]{Klein2018}
Klein,~D.~R.; MacNeill,~D.; Lado,~J.~L.; Soriano,~D.; Navarro-Moratalla,~E.; Watanabe,~K.; Taniguchi,~T.; Manni,~S.; Canfield,~P.; Fern{\'{a}}ndez-Rossier,~J.; Jarillo-Herrero,~P. {Probing magnetism in 2D van der Waals crystalline insulators via electron tunneling}. \emph{Science} \textbf{2018}, \emph{360}, 1218--1222\relax
\mciteBstWouldAddEndPuncttrue
\mciteSetBstMidEndSepPunct{\mcitedefaultmidpunct}
{\mcitedefaultendpunct}{\mcitedefaultseppunct}\relax
\EndOfBibitem
\bibitem[Wang \latin{et~al.}(2021)Wang, Guti{\'{e}}rrez-Lezama, Dumcenco, Ubrig, Taniguchi, Watanabe, Giannini, Gibertini, and Morpurgo]{Wang2021}
Wang,~Z.; Guti{\'{e}}rrez-Lezama,~I.; Dumcenco,~D.; Ubrig,~N.; Taniguchi,~T.; Watanabe,~K.; Giannini,~E.; Gibertini,~M.; Morpurgo,~A.~F. {Magnetization dependent tunneling conductance of ferromagnetic barriers}. \emph{Nature Communications} \textbf{2021}, \emph{12}, 6659\relax
\mciteBstWouldAddEndPuncttrue
\mciteSetBstMidEndSepPunct{\mcitedefaultmidpunct}
{\mcitedefaultendpunct}{\mcitedefaultseppunct}\relax
\EndOfBibitem
\bibitem[Song \latin{et~al.}(2018)Song, Cai, Tu, Zhang, Huang, Wilson, Seyler, Zhu, Taniguchi, Watanabe, McGuire, Cobden, Xiao, Yao, and Xu]{Song2018}
Song,~T.; Cai,~X.; Tu,~M. W.~Y.; Zhang,~X.; Huang,~B.; Wilson,~N.~P.; Seyler,~K.~L.; Zhu,~L.; Taniguchi,~T.; Watanabe,~K.; McGuire,~M.~A.; Cobden,~D.~H.; Xiao,~D.; Yao,~W.; Xu,~X. {Giant tunneling magnetoresistance in spin-filter van der Waals heterostructures}. \emph{Science} \textbf{2018}, \emph{360}, 1214--1218\relax
\mciteBstWouldAddEndPuncttrue
\mciteSetBstMidEndSepPunct{\mcitedefaultmidpunct}
{\mcitedefaultendpunct}{\mcitedefaultseppunct}\relax
\EndOfBibitem
\bibitem[Wang \latin{et~al.}(2018)Wang, Guti{\'{e}}rrez-Lezama, Ubrig, Kroner, Gibertini, Taniguchi, Watanabe, Imamoǧlu, Giannini, and Morpurgo]{Wang2018}
Wang,~Z.; Guti{\'{e}}rrez-Lezama,~I.; Ubrig,~N.; Kroner,~M.; Gibertini,~M.; Taniguchi,~T.; Watanabe,~K.; Imamoǧlu,~A.; Giannini,~E.; Morpurgo,~A.~F. {Very large tunneling magnetoresistance in layered magnetic semiconductor CrI$_3$}. \emph{Nature Communications} \textbf{2018}, \emph{9}, 1--8\relax
\mciteBstWouldAddEndPuncttrue
\mciteSetBstMidEndSepPunct{\mcitedefaultmidpunct}
{\mcitedefaultendpunct}{\mcitedefaultseppunct}\relax
\EndOfBibitem
\bibitem[Cai \latin{et~al.}(2019)Cai, Song, Wilson, Clark, He, Zhang, Taniguchi, Watanabe, Yao, Xiao, McGuire, Cobden, and Xu]{Cai2019}
Cai,~X.; Song,~T.; Wilson,~N.~P.; Clark,~G.; He,~M.; Zhang,~X.; Taniguchi,~T.; Watanabe,~K.; Yao,~W.; Xiao,~D.; McGuire,~M.~A.; Cobden,~D.~H.; Xu,~X. {Atomically Thin CrCl$_3$: An In-Plane Layered Antiferromagnetic Insulator}. \emph{Nano Letters} \textbf{2019}, \emph{19}, 3993--3998\relax
\mciteBstWouldAddEndPuncttrue
\mciteSetBstMidEndSepPunct{\mcitedefaultmidpunct}
{\mcitedefaultendpunct}{\mcitedefaultseppunct}\relax
\EndOfBibitem
\bibitem[Long \latin{et~al.}(2020)Long, Henck, Gibertini, Dumcenco, Wang, Taniguchi, Watanabe, Giannini, and Morpurgo]{Long2020}
Long,~G.; Henck,~H.; Gibertini,~M.; Dumcenco,~D.; Wang,~Z.; Taniguchi,~T.; Watanabe,~K.; Giannini,~E.; Morpurgo,~A.~F. {Persistence of Magnetism in Atomically Thin MnPS$_3$ Crystals}. \emph{Nano Letters} \textbf{2020}, \emph{20}, 2452--2459\relax
\mciteBstWouldAddEndPuncttrue
\mciteSetBstMidEndSepPunct{\mcitedefaultmidpunct}
{\mcitedefaultendpunct}{\mcitedefaultseppunct}\relax
\EndOfBibitem
\bibitem[Kim \latin{et~al.}(2019)Kim, Yang, Li, Jiang, Jin, Tao, Nichols, Sfigakis, Zhong, Li, Tian, Cory, Miao, Shan, Mak, Lei, Sun, Zhao, and Tsen]{Kim2019a}
Kim,~H.~H. \latin{et~al.}  {Evolution of interlayer and intralayer magnetism in three atomically thin chromium trihalides}. \emph{Proceedings of the National Academy of Sciences of the United States of America} \textbf{2019}, \emph{166}, 11131--11136\relax
\mciteBstWouldAddEndPuncttrue
\mciteSetBstMidEndSepPunct{\mcitedefaultmidpunct}
{\mcitedefaultendpunct}{\mcitedefaultseppunct}\relax
\EndOfBibitem
\bibitem[Kim \latin{et~al.}(2018)Kim, Yang, Patel, Sfigakis, Li, Tian, Lei, and Tsen]{Kim2018}
Kim,~H.~H.; Yang,~B.; Patel,~T.; Sfigakis,~F.; Li,~C.; Tian,~S.; Lei,~H.; Tsen,~A.~W. {One Million Percent Tunnel Magnetoresistance in a Magnetic van der Waals Heterostructure}. \emph{Nano Letters} \textbf{2018}, \emph{18}, 4885--4890\relax
\mciteBstWouldAddEndPuncttrue
\mciteSetBstMidEndSepPunct{\mcitedefaultmidpunct}
{\mcitedefaultendpunct}{\mcitedefaultseppunct}\relax
\EndOfBibitem
\bibitem[Kim \latin{et~al.}(2019)Kim, Yang, Tian, Li, Miao, Lei, and Tsen]{Kim2019}
Kim,~H.~H.; Yang,~B.; Tian,~S.; Li,~C.; Miao,~G.~X.; Lei,~H.; Tsen,~A.~W. {Tailored Tunnel Magnetoresistance Response in Three Ultrathin Chromium Trihalides}. \emph{Nano Letters} \textbf{2019}, \emph{19}, 5739--5745\relax
\mciteBstWouldAddEndPuncttrue
\mciteSetBstMidEndSepPunct{\mcitedefaultmidpunct}
{\mcitedefaultendpunct}{\mcitedefaultseppunct}\relax
\EndOfBibitem
\bibitem[Klein \latin{et~al.}(2019)Klein, MacNeill, Song, Larson, Fang, Xu, Ribeiro, Canfield, Kaxiras, Comin, and Jarillo-Herrero]{Klein2019}
Klein,~D.~R.; MacNeill,~D.; Song,~Q.; Larson,~D.~T.; Fang,~S.; Xu,~M.; Ribeiro,~R.~A.; Canfield,~P.~C.; Kaxiras,~E.; Comin,~R.; Jarillo-Herrero,~P. {Enhancement of interlayer exchange in an ultrathin two-dimensional magnet}. \emph{Nature Physics} \textbf{2019}, \emph{15}, 1255--1260\relax
\mciteBstWouldAddEndPuncttrue
\mciteSetBstMidEndSepPunct{\mcitedefaultmidpunct}
{\mcitedefaultendpunct}{\mcitedefaultseppunct}\relax
\EndOfBibitem
\bibitem[Simmons(1963)]{Simmons1963}
Simmons,~J.~G. {Generalized Formula for the Electric Tunnel Effect between Similar Electrodes Separated by a Thin Insulating Film}. \emph{Journal of Applied Physics} \textbf{1963}, \emph{34}, 1793--1803\relax
\mciteBstWouldAddEndPuncttrue
\mciteSetBstMidEndSepPunct{\mcitedefaultmidpunct}
{\mcitedefaultendpunct}{\mcitedefaultseppunct}\relax
\EndOfBibitem
\bibitem[Esaki \latin{et~al.}(1967)Esaki, Stiles, and von Molnar]{Esaki1967}
Esaki,~L.; Stiles,~P.~J.; von Molnar,~S. {Magnetointernal field emission in junctions of magnetic insulators}. \emph{Physical Review Letters} \textbf{1967}, \emph{19}, 852--854\relax
\mciteBstWouldAddEndPuncttrue
\mciteSetBstMidEndSepPunct{\mcitedefaultmidpunct}
{\mcitedefaultendpunct}{\mcitedefaultseppunct}\relax
\EndOfBibitem
\bibitem[Gopinadhan \latin{et~al.}(2015)Gopinadhan, Shin, Jalil, Venkatesan, Geim, Neto, and Yang]{Gopinadhan2015}
Gopinadhan,~K.; Shin,~Y.~J.; Jalil,~R.; Venkatesan,~T.; Geim,~A.~K.; Neto,~A.~H.; Yang,~H. {Extremely large magnetoresistance in few-layer graphene/boron-nitride heterostructures}. \emph{Nature Communications} \textbf{2015}, \emph{6}, 1--7\relax
\mciteBstWouldAddEndPuncttrue
\mciteSetBstMidEndSepPunct{\mcitedefaultmidpunct}
{\mcitedefaultendpunct}{\mcitedefaultseppunct}\relax
\EndOfBibitem
\bibitem[Matos-Abiague and Fabian(2009)Matos-Abiague, and Fabian]{Matos-Abiague2009}
Matos-Abiague,~A.; Fabian,~J. {Anisotropic tunneling magnetoresistance and tunneling anisotropic magnetoresistance: Spin-orbit coupling in magnetic tunnel junctions}. \emph{Physical Review B} \textbf{2009}, \emph{79}, 155303\relax
\mciteBstWouldAddEndPuncttrue
\mciteSetBstMidEndSepPunct{\mcitedefaultmidpunct}
{\mcitedefaultendpunct}{\mcitedefaultseppunct}\relax
\EndOfBibitem
\bibitem[Tang \latin{et~al.}(2022)Tang, Wang, and Jia]{Tang2022}
Tang,~H.-M.; Wang,~S.-Z.; Jia,~X.-T. {Tunneling anisotropic magnetoresistance in MgO-based magnetic tunnel junctions induced by spin-orbit coupling}. \emph{Physical Review B} \textbf{2022}, \emph{106}, 094406\relax
\mciteBstWouldAddEndPuncttrue
\mciteSetBstMidEndSepPunct{\mcitedefaultmidpunct}
{\mcitedefaultendpunct}{\mcitedefaultseppunct}\relax
\EndOfBibitem
\bibitem[Sears \latin{et~al.}(2020)Sears, Chern, Kim, Bereciartua, Francoual, Kim, and Kim]{Sears2020}
Sears,~J.~A.; Chern,~L.~E.; Kim,~S.; Bereciartua,~P.~J.; Francoual,~S.; Kim,~Y.~B.; Kim,~Y.~J. {Ferromagnetic Kitaev interaction and the origin of large magnetic anisotropy in $\alpha$-RuCl$_3$}. \emph{Nature Physics} \textbf{2020}, \emph{16}, 837--840\relax
\mciteBstWouldAddEndPuncttrue
\mciteSetBstMidEndSepPunct{\mcitedefaultmidpunct}
{\mcitedefaultendpunct}{\mcitedefaultseppunct}\relax
\EndOfBibitem
\bibitem[Rau \latin{et~al.}(2014)Rau, Lee, and Kee]{Rau2014}
Rau,~J.~G.; Lee,~E. K.-H.; Kee,~H.-Y. {Generic Spin Model for the Honeycomb Iridates beyond the Kitaev Limit}. \emph{Physical Review Letters} \textbf{2014}, \emph{112}, 077204\relax
\mciteBstWouldAddEndPuncttrue
\mciteSetBstMidEndSepPunct{\mcitedefaultmidpunct}
{\mcitedefaultendpunct}{\mcitedefaultseppunct}\relax
\EndOfBibitem
\bibitem[Chaloupka and Khaliullin(2016)Chaloupka, and Khaliullin]{Chaloupka2016}
Chaloupka,~J.; Khaliullin,~G. {Magnetic anisotropy in the Kitaev model systems Na$_2$IrO$_3$ and RuCl$_3$}. \emph{Physical Review B} \textbf{2016}, \emph{94}, 64435\relax
\mciteBstWouldAddEndPuncttrue
\mciteSetBstMidEndSepPunct{\mcitedefaultmidpunct}
{\mcitedefaultendpunct}{\mcitedefaultseppunct}\relax
\EndOfBibitem
\bibitem[Kubota \latin{et~al.}(2015)Kubota, Tanaka, Ono, Narumi, and Kindo]{Kubota2015}
Kubota,~Y.; Tanaka,~H.; Ono,~T.; Narumi,~Y.; Kindo,~K. {Successive magnetic phase transitions in $\alpha$-RuCl$_3$: XY-like frustrated magnet on the honeycomb lattice}. \emph{Physical Review B} \textbf{2015}, \emph{91}, 094422\relax
\mciteBstWouldAddEndPuncttrue
\mciteSetBstMidEndSepPunct{\mcitedefaultmidpunct}
{\mcitedefaultendpunct}{\mcitedefaultseppunct}\relax
\EndOfBibitem
\bibitem[Cen and Kee(2022)Cen, and Kee]{Cen2022}
Cen,~J.; Kee,~H.-Y. {Strategy to extract Kitaev interaction using symmetry in honeycomb Mott insulators}. \emph{Communications Physics} \textbf{2022}, \emph{5}, 119\relax
\mciteBstWouldAddEndPuncttrue
\mciteSetBstMidEndSepPunct{\mcitedefaultmidpunct}
{\mcitedefaultendpunct}{\mcitedefaultseppunct}\relax
\EndOfBibitem
\bibitem[Tanaka \latin{et~al.}(2022)Tanaka, Mizukami, Harasawa, Hashimoto, Hwang, Kurita, Tanaka, Fujimoto, Matsuda, Moon, and Shibauchi]{Tanaka2022}
Tanaka,~O.; Mizukami,~Y.; Harasawa,~R.; Hashimoto,~K.; Hwang,~K.; Kurita,~N.; Tanaka,~H.; Fujimoto,~S.; Matsuda,~Y.; Moon,~E.~G.; Shibauchi,~T. {Thermodynamic evidence for a field-angle-dependent Majorana gap in a Kitaev spin liquid}. \emph{Nature Physics} \textbf{2022}, \emph{18}, 429--435\relax
\mciteBstWouldAddEndPuncttrue
\mciteSetBstMidEndSepPunct{\mcitedefaultmidpunct}
{\mcitedefaultendpunct}{\mcitedefaultseppunct}\relax
\EndOfBibitem
\bibitem[Balz \latin{et~al.}(2021)Balz, Janssen, Lampen-Kelley, Banerjee, Liu, Yan, Mandrus, Vojta, and Nagler]{Balz2021}
Balz,~C.; Janssen,~L.; Lampen-Kelley,~P.; Banerjee,~A.; Liu,~Y.~H.; Yan,~J.~Q.; Mandrus,~D.~G.; Vojta,~M.; Nagler,~S.~E. {Field-induced intermediate ordered phase and anisotropic interlayer interactions in $\alpha$-RuCl$_3$}. \emph{Physical Review B} \textbf{2021}, \emph{103}, 174417\relax
\mciteBstWouldAddEndPuncttrue
\mciteSetBstMidEndSepPunct{\mcitedefaultmidpunct}
{\mcitedefaultendpunct}{\mcitedefaultseppunct}\relax
\EndOfBibitem
\bibitem[Lampen-Kelley \latin{et~al.}(2018)Lampen-Kelley, Rachel, Reuther, Yan, Banerjee, Bridges, Cao, Nagler, and Mandrus]{Lampen-Kelley2018}
Lampen-Kelley,~P.; Rachel,~S.; Reuther,~J.; Yan,~J.~Q.; Banerjee,~A.; Bridges,~C.~A.; Cao,~H.~B.; Nagler,~S.~E.; Mandrus,~D. {Anisotropic susceptibilities in the honeycomb Kitaev system $\alpha$-RuCl$_3$}. \emph{Physical Review B} \textbf{2018}, \emph{98}, 100403\relax
\mciteBstWouldAddEndPuncttrue
\mciteSetBstMidEndSepPunct{\mcitedefaultmidpunct}
{\mcitedefaultendpunct}{\mcitedefaultseppunct}\relax
\EndOfBibitem
\bibitem[Guo \latin{et~al.}(2016)Guo, Liu, Yin, Wei, Lin, Hoffman, Zhao, Edgar, Chen, Lau, Dai, Yao, Wong, and Chai]{Guo2016}
Guo,~Y.; Liu,~C.; Yin,~Q.; Wei,~C.; Lin,~S.; Hoffman,~T.~B.; Zhao,~Y.; Edgar,~J.~H.; Chen,~Q.; Lau,~S.~P.; Dai,~J.; Yao,~H.; Wong,~H.~S.; Chai,~Y. {Distinctive in-Plane Cleavage Behaviors of Two-Dimensional Layered Materials}. \emph{ACS Nano} \textbf{2016}, \emph{10}, 8980--8988\relax
\mciteBstWouldAddEndPuncttrue
\mciteSetBstMidEndSepPunct{\mcitedefaultmidpunct}
{\mcitedefaultendpunct}{\mcitedefaultseppunct}\relax
\EndOfBibitem
\bibitem[Cao \latin{et~al.}(2016)Cao, Banerjee, Yan, Bridges, Lumsden, Mandrus, Tennant, Chakoumakos, and Nagler]{Cao2016}
Cao,~H.~B.; Banerjee,~A.; Yan,~J.~Q.; Bridges,~C.~A.; Lumsden,~M.~D.; Mandrus,~D.~G.; Tennant,~D.~A.; Chakoumakos,~B.~C.; Nagler,~S.~E. {Low-temperature crystal and magnetic structure of $\alpha$-RuCl$_3$}. \emph{Physical Review B} \textbf{2016}, \emph{93}, 134423\relax
\mciteBstWouldAddEndPuncttrue
\mciteSetBstMidEndSepPunct{\mcitedefaultmidpunct}
{\mcitedefaultendpunct}{\mcitedefaultseppunct}\relax
\EndOfBibitem
\bibitem[Sahasrabudhe \latin{et~al.}(2020)Sahasrabudhe, Kaib, Reschke, German, Koethe, Buhot, Kamenskyi, Hickey, Becker, Tsurkan, Loidl, Do, Choi, Gr{\"{u}}ninger, Winter, Wang, Valent{\'{i}}, and {Van Loosdrecht}]{Sahasrabudhe2020}
Sahasrabudhe,~A. \latin{et~al.}  {High-field quantum disordered state in $\alpha$-RuCl$_3$: Spin flips, bound states, and multiparticle continuum}. \emph{Physical Review B} \textbf{2020}, \emph{101}, 140410--140411\relax
\mciteBstWouldAddEndPuncttrue
\mciteSetBstMidEndSepPunct{\mcitedefaultmidpunct}
{\mcitedefaultendpunct}{\mcitedefaultseppunct}\relax
\EndOfBibitem
\bibitem[Ponomaryov \latin{et~al.}(2020)Ponomaryov, Zviagina, Wosnitza, Lampen-Kelley, Banerjee, Yan, Bridges, Mandrus, Nagler, and Zvyagin]{Ponomaryov2020}
Ponomaryov,~A.~N.; Zviagina,~L.; Wosnitza,~J.; Lampen-Kelley,~P.; Banerjee,~A.; Yan,~J.~Q.; Bridges,~C.~A.; Mandrus,~D.~G.; Nagler,~S.~E.; Zvyagin,~S.~A. {Nature of Magnetic Excitations in the High-Field Phase of $\alpha$-RuCl$_3$}. \emph{Physical Review Letters} \textbf{2020}, \emph{125}, 037202\relax
\mciteBstWouldAddEndPuncttrue
\mciteSetBstMidEndSepPunct{\mcitedefaultmidpunct}
{\mcitedefaultendpunct}{\mcitedefaultseppunct}\relax
\EndOfBibitem
\bibitem[Mi \latin{et~al.}(2021)Mi, Wang, Gui, Pi, Zheng, Yang, Gan, Wang, Li, Wang, Zhang, Su, Chai, and He]{Mi2021}
Mi,~X.; Wang,~X.; Gui,~H.; Pi,~M.; Zheng,~T.; Yang,~K.; Gan,~Y.; Wang,~P.; Li,~A.; Wang,~A.; Zhang,~L.; Su,~Y.; Chai,~Y.; He,~M. {Stacking faults in $\alpha$-RuCl$_3$ revealed by local electric polarization}. \emph{Physical Review B} \textbf{2021}, \emph{103}, 174413\relax
\mciteBstWouldAddEndPuncttrue
\mciteSetBstMidEndSepPunct{\mcitedefaultmidpunct}
{\mcitedefaultendpunct}{\mcitedefaultseppunct}\relax
\EndOfBibitem
\bibitem[Kim and Kee(2016)Kim, and Kee]{Kim2016}
Kim,~H.~S.; Kee,~H.~Y. {Crystal structure and magnetism in $\alpha$-RuCl$_3$: An ab initio study}. \emph{Physical Review B} \textbf{2016}, \emph{93}, 155143\relax
\mciteBstWouldAddEndPuncttrue
\mciteSetBstMidEndSepPunct{\mcitedefaultmidpunct}
{\mcitedefaultendpunct}{\mcitedefaultseppunct}\relax
\EndOfBibitem
\bibitem[Hart \latin{et~al.}(2023)Hart, Bhatt, Zhu, Han, Bianco, Li, Hynek, Schneeloch, Tao, Louca, Guo, Zhu, Jornada, Reed, Kourkoutis, and Cha]{Hart2023}
Hart,~J.~L. \latin{et~al.}  {Emergent layer stacking arrangements in c-axis confined MoTe$_2$}. \emph{Nature Communications} \textbf{2023}, \emph{14}, 4803\relax
\mciteBstWouldAddEndPuncttrue
\mciteSetBstMidEndSepPunct{\mcitedefaultmidpunct}
{\mcitedefaultendpunct}{\mcitedefaultseppunct}\relax
\EndOfBibitem
\bibitem[Hovden \latin{et~al.}(2016)Hovden, Tsen, Liu, Savitzky, {El Baggari}, Liu, Lu, Sun, Kim, Pasupathy, and Kourkoutis]{Hovden2016}
Hovden,~R.; Tsen,~A.~W.; Liu,~P.; Savitzky,~B.~H.; {El Baggari},~I.; Liu,~Y.; Lu,~W.; Sun,~Y.; Kim,~P.; Pasupathy,~A.~N.; Kourkoutis,~L.~F. {Atomic lattice disorder in charge-density-wave phases of exfoliated dichalcogenides (1T-TaS$_2$)}. \emph{Proceedings of the National Academy of Sciences of the United States of America} \textbf{2016}, \emph{113}, 11420--11424\relax
\mciteBstWouldAddEndPuncttrue
\mciteSetBstMidEndSepPunct{\mcitedefaultmidpunct}
{\mcitedefaultendpunct}{\mcitedefaultseppunct}\relax
\EndOfBibitem
\bibitem[Reschke \latin{et~al.}(2017)Reschke, Mayr, Wang, Do, Choi, and Loidl]{Reschke2017}
Reschke,~S.; Mayr,~F.; Wang,~Z.; Do,~S.~H.; Choi,~K.~Y.; Loidl,~A. {Electronic and phonon excitations in $\alpha$-RuCl$_3$}. \emph{Physical Review B} \textbf{2017}, \emph{96}, 165120\relax
\mciteBstWouldAddEndPuncttrue
\mciteSetBstMidEndSepPunct{\mcitedefaultmidpunct}
{\mcitedefaultendpunct}{\mcitedefaultseppunct}\relax
\EndOfBibitem
\bibitem[Glamazda \latin{et~al.}(2017)Glamazda, Lemmens, Do, Kwon, and Choi]{Glamazda2017}
Glamazda,~A.; Lemmens,~P.; Do,~S.~H.; Kwon,~Y.~S.; Choi,~K.~Y. {Relation between Kitaev magnetism and structure in $\alpha$-RuCl 3}. \emph{Physical Review B} \textbf{2017}, \emph{95}, 174429\relax
\mciteBstWouldAddEndPuncttrue
\mciteSetBstMidEndSepPunct{\mcitedefaultmidpunct}
{\mcitedefaultendpunct}{\mcitedefaultseppunct}\relax
\EndOfBibitem
\bibitem[He \latin{et~al.}(2018)He, Zhong, Kim, Ye, Ye, Winford, McHaffie, Rilak, Chen, Luo, Sun, and Tsen]{He2018}
He,~R.; Zhong,~S.; Kim,~H.~H.; Ye,~G.; Ye,~Z.; Winford,~L.; McHaffie,~D.; Rilak,~I.; Chen,~F.; Luo,~X.; Sun,~Y.; Tsen,~A.~W. {Dimensionality-driven orthorhombic MoTe$_2$ at room temperature}. \emph{Physical Review B} \textbf{2018}, \emph{97}, 041410\relax
\mciteBstWouldAddEndPuncttrue
\mciteSetBstMidEndSepPunct{\mcitedefaultmidpunct}
{\mcitedefaultendpunct}{\mcitedefaultseppunct}\relax
\EndOfBibitem
\bibitem[Li \latin{et~al.}(2019)Li, Jiang, Sivadas, Wang, Xu, Weber, Goldberger, Watanabe, Taniguchi, Fennie, {Fai Mak}, and Shan]{Li2019}
Li,~T.; Jiang,~S.; Sivadas,~N.; Wang,~Z.; Xu,~Y.; Weber,~D.; Goldberger,~J.~E.; Watanabe,~K.; Taniguchi,~T.; Fennie,~C.~J.; {Fai Mak},~K.; Shan,~J. {Pressure-controlled interlayer magnetism in atomically thin CrI$_3$}. \emph{Nature Materials} \textbf{2019}, \emph{18}, 1303--1308\relax
\mciteBstWouldAddEndPuncttrue
\mciteSetBstMidEndSepPunct{\mcitedefaultmidpunct}
{\mcitedefaultendpunct}{\mcitedefaultseppunct}\relax
\EndOfBibitem
\bibitem[Song \latin{et~al.}(2019)Song, Fei, Yankowitz, Lin, Jiang, Hwangbo, Zhang, Sun, Taniguchi, Watanabe, McGuire, Graf, Cao, Chu, Cobden, Dean, Xiao, and Xu]{Song2019}
Song,~T. \latin{et~al.}  {Switching 2D magnetic states via pressure tuning of layer stacking}. \emph{Nature Materials} \textbf{2019}, \emph{18}, 1298--1302\relax
\mciteBstWouldAddEndPuncttrue
\mciteSetBstMidEndSepPunct{\mcitedefaultmidpunct}
{\mcitedefaultendpunct}{\mcitedefaultseppunct}\relax
\EndOfBibitem
\bibitem[Song \latin{et~al.}(2021)Song, Sun, Anderson, Wang, Qian, Taniguchi, Watanabe, McGuire, St{\"{o}}hr, Xiao, Cao, Wrachtrup, and Xu]{Song2021}
Song,~T.; Sun,~Q.~C.; Anderson,~E.; Wang,~C.; Qian,~J.; Taniguchi,~T.; Watanabe,~K.; McGuire,~M.~A.; St{\"{o}}hr,~R.; Xiao,~D.; Cao,~T.; Wrachtrup,~J.; Xu,~X. {Direct visualization of magnetic domains and moir{\'{e}} magnetism in twisted 2D magnets}. \emph{Science} \textbf{2021}, \emph{374}, 1140--1144\relax
\mciteBstWouldAddEndPuncttrue
\mciteSetBstMidEndSepPunct{\mcitedefaultmidpunct}
{\mcitedefaultendpunct}{\mcitedefaultseppunct}\relax
\EndOfBibitem
\bibitem[Kresse and Hafner(1993)Kresse, and Hafner]{Kresse1993}
Kresse,~G.; Hafner,~J. {Ab initio molecular dynamics for liquid metals}. \emph{Physical Review B} \textbf{1993}, \emph{47}, 558--561\relax
\mciteBstWouldAddEndPuncttrue
\mciteSetBstMidEndSepPunct{\mcitedefaultmidpunct}
{\mcitedefaultendpunct}{\mcitedefaultseppunct}\relax
\EndOfBibitem
\bibitem[Bl{\"{o}}chl(1994)]{Blochl1994}
Bl{\"{o}}chl,~P.~E. {Projector augmented-wave method}. \emph{Physical Review B} \textbf{1994}, \emph{50}, 17953--17979\relax
\mciteBstWouldAddEndPuncttrue
\mciteSetBstMidEndSepPunct{\mcitedefaultmidpunct}
{\mcitedefaultendpunct}{\mcitedefaultseppunct}\relax
\EndOfBibitem
\bibitem[Perdew \latin{et~al.}(1996)Perdew, Burke, and Ernzerhof]{Perdew1996}
Perdew,~J.~P.; Burke,~K.; Ernzerhof,~M. {Generalized Gradient Approximation Made Simple}. \emph{Physical Review Letters} \textbf{1996}, \emph{77}, 3865--3868\relax
\mciteBstWouldAddEndPuncttrue
\mciteSetBstMidEndSepPunct{\mcitedefaultmidpunct}
{\mcitedefaultendpunct}{\mcitedefaultseppunct}\relax
\EndOfBibitem
\bibitem[Kim \latin{et~al.}(2015)Kim, V., Catuneanu, and Kee]{Kim2015}
Kim,~H.-S.; V.,~V.~S.; Catuneanu,~A.; Kee,~H.-Y. {Kitaev magnetism in honeycomb RuCl$_3$ with intermediate spin-orbit coupling}. \emph{Physical Review B} \textbf{2015}, \emph{91}, 241110\relax
\mciteBstWouldAddEndPuncttrue
\mciteSetBstMidEndSepPunct{\mcitedefaultmidpunct}
{\mcitedefaultendpunct}{\mcitedefaultseppunct}\relax
\EndOfBibitem
\end{mcitethebibliography}

\end{document}